\documentclass[a4paper,11pt]{article}
\usepackage{setspace}
\usepackage[top=2.54cm,bottom=2.54cm, left=2.54cm, right=2.54cm]{geometry}
\usepackage{bm}
\usepackage{natbib}
\usepackage{graphicx}
\usepackage{epstopdf}
\usepackage{amsmath}
\usepackage{hyperref}
\usepackage{caption}
\usepackage{amsfonts}

\begin{document}
\bibliographystyle{apalike}

\begin{center}
\Large{Improved classification for compositional data using the $\alpha$-transformation}
\\
\normalsize
\bigskip \bigskip
{\bf Michail Tsagris$^1$, Simon Preston$^2$ and Andrew T.A. Wood$^2$} \\
$^1$ Department of Computer Science, University of Crete, Heraklion, Greece \\
$^2$ School of Mathematical Sciences, University of Nottingham, UK \\
\href{mailto:mtsagris@yahoo.gr}{mtsagris@yahoo.gr}, 
\href{mailto:Simon.Preston@nottingham.ac.uk}{Simon.Preston@nottingham.ac.uk} and
\href{mailto:Andrew.Wood@nottingham.ac.uk}{Andrew.Wood@nottingham.ac.uk}
\end{center}

\begin{center}
{\bf Abstract}
\end{center}
In compositional data analysis an observation is a vector containing non-negative values,
only the relative sizes of which are considered to be of interest.
Without loss of generality, a compositional vector can be taken to be a
vector of proportions that sum to one.  Data of this type arise in many areas including
geology, archaeology, biology, economics and political science.  In this paper we
investigate methods for classification of compositional data.  Our approach centres on the
idea of using the $\alpha$-transformation to transform the data and then to classify the
transformed data via regularised discriminant analysis and the $k$-nearest neighbours
algorithm.  Using the $\alpha$-transformation generalises two rival approaches in
compositional data analysis, one (when $\alpha$=1) that treats the data as though they
were Euclidean, ignoring the compositional constraint, and another (when $\alpha =
0$) that employs Aitchison's centred log-ratio transformation.  A numerical study with
several real datasets shows that whether using $\alpha=1$ or $\alpha = 0$ gives
better classification performance depends on the dataset, and moreover that using an
intermediate value of $\alpha$ can sometimes give better performance
than using either 1 or 0. 

\section{Introduction}

Compositional data arise commonly in many fields, for instance geology \citep{ait1982}, in
studying constitution of rock samples; economics \citep{fry2000}, in budget allocations;
archaeology \citep{baxter2005}, in the constitution of man-made glasses; and the political
sciences \citep{rodrigues2009}, in voting behaviour.  In compositional data analysis, a
composition is considered an equivalence class comprising the set of 
multivariate vectors that differ only by a scalar factor and have non-negative
components.  Consequently, without loss of
generality, an observation may be viewed as a vector of proportions, i.e., with
non-negative components constrained to sum to 1.  The sample space of the observations is
hence the simplex 
\begin{eqnarray*} \mathbb{S}^d=\left\lbrace(x_1,...,x_D)^T
  \bigg\vert x_i \geq 0,\sum_{i=1}^Dx_i=1\right\rbrace, 
\end{eqnarray*} where $D$ denotes
  the number of components of the vector and $d=D-1$.

For statistical analysis of compositional data the question of how to account for the
compositional constraint arises.  A simple approach is to ignore the compositional
constraint and treat the data as though they were Euclidean, an approach we will call
``Euclidean data analysis'' (EDA) \citep{baxter2001, baxter2005,baxter2006, woronow1997}.
There is a school of thought, however, largely following from the work of \citet{ait1982,
ait1983, ait1992}, that ignoring the compositional constraint is inappropriate and can
lead to misleading inferences.  Aitchison contended that data should instead be analysed
after applying a ``logratio'' transformation, arguing that this amounted to working with
an implied distance measure on the simplex (discussed further in the next section) that
satisfied particular mathematical properties he regarded as essential for compositional
data analysis.  Other approaches to compositional data analysis that we
mention here but 
do not consider further in this paper include using different transformations, such as the
square-root transformation \citep{stephens1982, scealy2011b}, and parametric modelling,
for example using the Dirichlet distribution \citep{gueorguieva2008}.

Both EDA and Aitchison's logratio analysis (LRA) approach are widely used and there has
been a long and ongoing disagreement over which of these approaches, or indeed
others, is most appropriate to use.  The debate remains largely centred on the distance
measures implied by the various approaches and whether or not they satisfy particular
mathematical properties.  
\citet{scealy2014colours} have recently presented a historical
summary of the debate, and have given a critical appraisal of the properties often invoked
by authors to support the use of LRA.  We share Scealy and Welsh's opinion that LRA
should not be a default choice for compositional data analysis on account of such
properties.  In this paper, we take the pragmatic view, which seems especially relevant
for classification problems (in which out-of-sample classification error rate provides
an objective measure of performance), that we should adopt whichever approach performs
best in a given setting.

Indeed, a key message of this paper is that for classification problems, the choice of whether
or not one should transform the data, and if so which transformation to use, should
depend on the dataset under study.  This conclusion is clear from the fact that we can
easily generate a synthetic dataset for which LRA will perform perfectly and EDA poorly,
and vice versa.

One characteristic of a dataset that immediately rules out using LRA in its standard form
is the presence of observations for which one or more components is zero, since for such
observations the logratio transformation is undefined.  Data of this type are not uncommon
(in \S\ref{sec:applications} we consider two datasets containing observations with zeros),
so this is a notable weakness of LRA.  Some attempts have been made to modify LRA to make
it appropriate for data containing zeros (particularly when the zeros are assumed to arise
from rounding error), but these involve a somewhat ad hoc imputation approach of replacing
zeros with small values.  On a different tack, \citet{butler2008} developed parametric
models specifically for compositional data with zeros.  

Perhaps due to the nature of the data, little attention has been given to the problem of
classifying compositional data, especially where zero values are present.  Exceptions
are \citet{zadora2010} and \citet{neocleous2011}, who consider classification using
parametric models to account for the possibility of zeros values; see also
\citet{palarea2012} who consider the related problem of cluster analysis.   Our goal in
this paper is to develop adaptive classification algorithms which take into account the
characteristics of individual datasets, such as the distribution of the groups and the presence of zeros.  The main idea is to employ the Box--Cox-type $\alpha$-transformation
explored in \citet{tsagris2011}, and then use the transformed data as a basis for
classification.  This transformation has a free parameter, $\alpha$, and is such that the
case $\alpha = 0$ corresponds to the logratio transformation, and $\alpha=1$ corresponds
to a linear transformation of the data.  Hence using $\alpha = 0$ corresponds to LRA, and
$\alpha=1$, when used in conjunction with the discriminant analysis and
nearest-neighbour classification algorithms that we consider in \S\ref{sec:techniques}, is
equivalent to EDA.  For values of $\alpha$ between 0 and 1, the $\alpha$-transformation
offers a compromise between LRA and EDA.  An important benefit of the
$\alpha$-transformation is that it is well-defined for any $\alpha > 0$ for
compositions containing zeros.

The paper is structured as follows. In \S2 we discuss in more detail the
$\alpha$-transformation and the logratio transformation, and the associated implied
distance measures, and then in \S3 we consider some classification techniques and
how their performance can be improved using the $\alpha$-transformation. In \S4 we
present the results of 
a numerical study with four real datasets to investigate the performance of the
various techniques.  We conclude in \S5 with a discussion of the results.

\section{The $\alpha$-transformation and implied simplicial distance measure}

The $\alpha$-transformation of a compositional 
vector $\mathbf{x} \in \mathbb{S}^d$ (see \citet{tsagris2011}) is defined by
\begin{eqnarray} \label{alpha}
  \mathbf{z}_{\alpha}\left({\bf x}\right)={\bf H} \cdot \left(\frac{D \,{\bf
  u}_\alpha(\mathbf{x})-{\bf 1}_D}{\alpha}\right),
\end{eqnarray}
with $\alpha>0$ (we discuss more general $\alpha$ below), and where
\begin{eqnarray} \label{stayalpha}
  {\bf u}_\alpha(\mathbf{x})=\left( \frac{x_1^{\alpha}}{\sum_{j=1}^Dx_j^{\alpha}}, \ldots, \frac{x_D^{\alpha}}{\sum_{j=1}^Dx_j^{\alpha}} \right)^T
\end{eqnarray}
is the compositional power transformation \citep{ait2003},  ${\bf 1}_D$ is the
$D$-dimensional vector of ones, and $\bf H$ is 
any $d$-by-$D$ matrix
consisting of orthonormal rows, each of which is orthogonal to ${\bf 1}_D$;
similar ideas have been used in the compositional data context by \citet{egoz2003} and, of
course, in many other contexts.  
A suitable choice for $\mathbf{H}$ (noting in any case that the classification methods
in this paper are invariant to the particular choice) is the Helmert matrix
\citep{helm1965, dryden1998ssa} with the first row removed, i.e., 
the matrix whose $j$th row is
\begin{equation}
  (h_j,\dots,h_j, -jh_j, 0, \dots, 0), \quad \text{where} \quad h_j = - \left\{ j(j+1)
  \right\}^{-1/2},
\end{equation}
with $h_j$ repeated $j$ times and $0$ repeated $d-j$ times.   The purpose of
$\mathbf{H}$ is to remove the redundant dimension which is present due to the
compositional constraint.  In particular, the vector $\left({D \,{\bf
u}_\alpha(\mathbf{x})-{\bf 1}_D}\right)/{\alpha}$ has components which sum to zero and
therefore it lies in a subspace of $\mathbb{R}^D$; left-multiplication by $\mathbf{H}$ is
an isometric one-to-one mapping from this subspace into $\mathbb{R}^d$.
The image $\mathcal{V}_\alpha = \left\{ \mathbf{z}_\alpha(\mathbf{x}) : \mathbf{x} \in
\mathbb{S}^d \right\}$
of transformation (\ref{alpha}) is $\mathbb{R}^d$ in the limit $\alpha \rightarrow 0$ but
a strict subset of $\mathbb{R}^d$ for $\alpha \neq 0$.  Transformation (\ref{alpha}) is
invertible: for $\mathbf{v} \in \mathcal{V}_\alpha$ the inverse of
$\mathbf{z}_\alpha(\mathbf{x})$ is
\begin{eqnarray}
  \mathbf{z}_{\alpha}^{-1}\left({\bf v}\right)=
  \mathbf{u}_\alpha^{-1} \left( \alpha \mathbf{H}^\top \mathbf{v} + \mathbf{1}_D \right)
  \in \mathbb{S}^d,
  \label{alphaInverse}
\end{eqnarray}
where 
\begin{eqnarray}
  {\bf u}_\alpha^{-1}(\mathbf{x})
=\left(\frac{x_1^{1/\alpha}}{\sum_{j=1}^Dx_j^{1/\alpha}},\ldots,\frac{x_D^{1/\alpha}}{\sum_{j=1}^Dx_j^{1/\alpha}}\right).
\end{eqnarray}
If one is willing to exclude from the sample space the boundary of the simplex, which
corresponds to observations that have one or more components equal to zero, then the
$\alpha$-transformation (\ref{alpha}) and its inverse (\ref{alphaInverse}) are well defined
for all $\alpha \in \mathbb{R}$.  (Excluding the
boundary is standard practise in LRA because the definition is used to sidestep 
the problem of having data with zeros.)  
The motivation for transformation (\ref{alpha}) is that the case $\alpha = 0$
corresponds to LRA, whereas $\alpha = 1$ corresponds to EDA.  We define the case $\alpha = 0$ in
terms of the limit $\alpha \rightarrow 0$; then 
\begin{equation}
  \mathbf{z}_0(\mathbf{x}) = \lim_{\alpha \rightarrow 0} \mathbf{z}_\alpha(\mathbf{x})
= \mathbf{H} \cdot \mathbf{w}(\mathbf{x}),
\label{ilr}
\end{equation}
where
\begin{eqnarray}
 {\bf w}(\mathbf{x})=\left( \log \left\{\frac{x_1}{g(\mathbf{x})} \right\}, \ldots, 
 \log \left\{\frac{x_D}{g(\mathbf{x})}\right\} \right)^T,
  \label{aitchisonclr}
\end{eqnarray}
is Aitchison's centred logratio transformation \citep{ait1983,ait2003} and $g\left({\bf x} \right)=\prod_{i=1}^Dx_i^{1/D}$ 
is the geometric mean of the components of $\mathbf{x}$. See the Appendix for 
proof of (\ref{ilr}).  For the case $\alpha = 1$,
(\ref{alpha}) is just a linear
transformation of the simplex. 

Power transformations similar to (\ref{alpha}) were
considered by \citet{greenacre2009b} and \citet{greenacre2011}, in the somewhat
different context of correspondence analysis.  A Box--Cox
transformation applied 
to each component of $\mathbf{x} \in \mathbb{S}^d$ 
so that $\mathbf{x}$ is transformed to
\begin{equation}
\left({\theta}^{-1} \left(x_1^{\theta}-1\right), \dots, {\theta}^{-1}
\left(x_D^{\theta}-1\right)\right)^T,
\end{equation}
has the limit $(\log x_1, \dots, \log x_D)^T$ as $\theta \rightarrow 0$.  We favour
transformation (\ref{alpha}) in this work in view of its closer connection, via
(\ref{ilr}), to Aitchison's centred logratio transformation.

The $\alpha$-transformation (\ref{alpha}) leads to a natural simplicial
distance measure $\Delta_{\alpha}\left({\bf x},{\bf y}\right)$, which we call the
$\alpha$-metric, between observations ${\bf
x},{\bf y} \in \mathbb{S}^d$, defined in terms of the Euclidean distance $\| \cdot \|$ between
transformed observations, i.e.,
\begin{align} 
\Delta_{\alpha}\left({\bf x},{\bf y}\right)
& = \| \mathbf{z}_\alpha(\mathbf{x}) - \mathbf{z}_\alpha(\mathbf{y}) \|
\nonumber
\\
& =\frac{D}{\left| \alpha \right|}\left[\sum_{i=1}^D\left(\frac{x_i^\alpha}
{\sum_{j=1}^D x_j^\alpha}-\frac{y_i^\alpha}{\sum_{j=1}^D
y_j^\alpha}\right)^2\right]^{1/2}.
\label{adist}
\end{align}
The special case
\begin{equation}
\Delta_0\left({\bf x},{\bf y}\right):=
\lim_{\alpha \rightarrow 0} \Delta_\alpha\left({\bf x},{\bf y}\right)
= \left[\sum_{i=1}^D\left(\log{\frac{x_i}{g\left({\bf x}\right)}}-\log{\frac{y_i}{g\left({\bf y}\right)}} \right)^2 \right]^{1/2},
\label{aitchisondistance}
\end{equation}
is Aitchison's distance measure \citep{ait2000}, whereas
\begin{eqnarray} \label{rawdist}
\Delta_1\left({\bf x},{\bf y}\right)=D\left[\sum_{i=1}^D\left(x_i-y_i\right)^2\right]^{1/2}
\end{eqnarray}
is just Euclidean distance multiplied by $D$.

Transformation (\ref{alpha}), and the implied distance measure (\ref{adist}), offer 
flexibility in data analysis: the choice of $\alpha$ enables either 
LRA or EDA, or a compromise between the two, and 
the particular value of $\alpha$ can be chosen to optimise some measure of
practical performance (in this paper, the out-of-sample classification error rate).
Crucially, for $\alpha > 0$, the transformation and distance are well defined even when
some components have zero values, in contrast to (\ref{aitchisonclr}) and
(\ref{aitchisondistance}). 

Amongst the criteria for compositional distance measures listed by \citet{ait1992}, the
distance measure (\ref{adist}) satisfies ``positivity''
($\Delta_\alpha(\mathbf{x},\mathbf{y})>0$ for $\mathbf{x} \neq \mathbf{y}$), ``zero
difference for equivalent compositions'' ($\Delta_\alpha(\mathbf{x}, \mathbf{x}) = 0)$,
``interchangeability of compositions'' ($\Delta_\alpha(\mathbf{x},\mathbf{y}) =
\Delta_\alpha(\mathbf{y},\mathbf{x})$), ``scale invariance'' ($\Delta_\alpha(a
\mathbf{x},A \mathbf{y}) = \Delta_\alpha(\mathbf{x},\mathbf{y})$ for all $a>0, \, A>0$)
and ``permutation invariance'' ($\Delta_\alpha(P \mathbf{x},P \mathbf{y}) =
\Delta_\alpha(\mathbf{x},\mathbf{y})$ for any permutation $P$).  It does not satisfy
``perturbation invariance'', a property strongly tied to the logratio transformation
\citep{ait2003}; and nor does it satisfy ``subcompositional coherence'', a criterion that
affects inferences regarding the relationships between compositional components
\citep{greenacre2011}.  The question of how much importance should be given to
subcompositional coherence in compositional data analysis has been a matter of much
debate; see for example the historical review and discussion in Scealy and Welsh (2014).
Our view is similar to that of Scealy and Welsh (2014), which is that subcompositional
dominance is not a property of primary importance, although we point out that a referee
strongly disagrees with our position.  We reiterate that our motivation is to achieve
strong practical performance, whether or not our distance measure satisfies any particular
properties.

\section{Classification techniques for compositional data}
\label{sec:techniques}

The key idea now is to use the $\alpha$-transformation (\ref{alpha}) in conjunction with
regularised descriminant analysis (RDA), and the $\alpha$-metric (\ref{adist}) in
conjunction with $k$-nearest-neighbours ($k$-NN) classification, to investigate how
performance for various values of $\alpha$ compares with the special cases of EDA
($\alpha=1$) and LRA ($\alpha = 0$).
We will begin with a brief review of regularised
discriminant analysis, of which linear and quadratic discriminant analysis are special
cases, and with the $k$-nearest neighbours algorithm. 

\subsection{Regularised discriminant analysis (RDA)}
In discriminant analysis we allocate an observation to the group with the
highest (posterior) density, assuming that observations in
each group come from a multivariate normal
distribution.  Given a training sample with $g$ groups containing $n_1, \ldots, n_g$
observations, then a new observation ${\bf z} \in \mathbb{R}^d$ is classified to the group
for which the discriminant score, $\delta_i(\mathbf{z})$, is largest, where 
\begin{eqnarray} \label{rule}
\delta_i\left({\bf z}\right)=-\frac{1}{2}\log{\left|2\pi\hat{\bm{\Sigma}}_i \right|}-
\frac{1}{2}\left({\bf z}-\hat{\bm{\mu}}_i\right)^T\hat{\bm{\Sigma}}_i^{-1}\left({\bf
z}-\hat{\bm{\mu}}_i\right)+\log{\pi_i};
\end{eqnarray}
here $|\cdot|$ denotes determinant, $\pi_i={n_i}/{n}$ with $n=\sum_{i=1}^gn_i$, and the $\hat{\bm{\mu}}_i$ and
$\hat{\bm{\Sigma}}_i$ are the sample mean vector and covariance matrix, respectively, of
the $i$th group.  Equation (\ref{rule}) is the Bayesian version of discriminant analysis,
incorporating the prior group membership probabilities
$\bm{\pi}=\left(\pi_1,\ldots,\pi_g\right)$, which assumes that the
proportions of observations in the training sample are representative of the proportions in
the population.  Other choices of $\bm{\pi}$ are possible depending on available prior
information.  The frequentist version uses instead $\pi_i = 1/g$.  We will use
the Bayesian version with $\pi_i={n_i}/{n}$ in our numerical investigations in $\S \ref{sec:applications}$.

The boundary between classification regions, say between groups $i$ and $j$, is defined by
$\delta_i(\mathbf{z}) = \delta_j(\mathbf{z})$.  From (\ref{rule}), the boundaries are
hence quadratic in $\mathbf{z}$, and for this reason the approach is termed
quadratic discriminant analysis (QDA). If we make the simplifying assumption that the groups
share a common covariance matrix, then the
$\hat{\bm{\Sigma}}_i$ in (\ref{rule})
can be replaced with the pooled estimate
\[
\hat{\bm{\Sigma}}_p=\frac{\sum_{i=1}^g\left(n_i-1\right)\hat{\bm{\Sigma}}_i}{n-g}.
\]
In this case, the boundaries are linear, and the approach is hence 
termed linear discriminant analysis
(LDA). 

QDA and LDA are special cases of so-called regularised discriminant
analysis (RDA); see \citet[pp.~112-113]{hastie2001}.
The idea of RDA is to regularise the covariance matrices by replacing them with 
weighted averages
\begin{eqnarray} \label{regcov}
\begin{array}{cccc}
& \hat{\bm{\Sigma}}_i\left(\lambda,\gamma \right) & = &
\lambda\hat{\bm{\Sigma}}_i+\left(1-\lambda\right)\hat{\bm{\Sigma}}_p\left(\gamma\right), \\
\text{and} & \hat{\bm{\Sigma}}_p\left(\gamma\right) & = &
\gamma\hat{\bm{\Sigma}}_p+\left(1-\gamma\right){\text{tr}\left(\hat{\bm{\Sigma}}_p\right)}{\bf
I}/d, \end{array} 
\end{eqnarray}
where $\lambda, \gamma \in [0,1]$ are two free parameters and $\mathbf{I}$ is the
$d$-by-$d$ identity matrix.  Parameter $\lambda$ offers a
trade-off between the more flexibile classification boundaries of QDA and the greater 
stability of LDA
to one or
more of the $\hat{\bm{\Sigma}}_i$ being ill-conditioned.  Parameter $\gamma$ offers further
stability if the pooled estimate $\hat{\bm{\Sigma}}_p$ is itself ill-conditioned.
Choosing $\lambda=1$ gives QDA, whereas choosing $\lambda=0$ and $\gamma=1$ gives LDA. 

We propose to use RDA with data transformed using the $\alpha$-transformation
(\ref{alpha}), and will denote this by RDA$\left(\alpha,\lambda,\gamma \right)$. Hence,
RDA$\left(0,\lambda,\gamma\right)$ amounts to the LRA approach of applying RDA 
to data transformed using the isometric log-ratio transformation (\ref{ilr}), 
whereas RDA$\left(1,\lambda,\gamma\right)$ amounts to the EDA approach of 
applying RDA to untransformed data.  We will also use the notation 
QDA($\alpha$) = RDA$\left(\alpha,1,0 \right)$ and 
LDA($\alpha$) = RDA$\left(\alpha,0,1 \right)$.  

\subsection{$k$-nearest neighbours ($k$-NN)}

The $k$-NN algorithm is an intuitive classifier that assumes no parametric model.  It
involves determining the $k$ observations in the training sample that are closest, by some
choice of distance measures, to the new test observation, then allocating the test
observation to the group most common amongst these $k$ ``nearest neighbours''.  Ties
caused by two or more groups jointly being most common can be broken by allocating
uniformly at random amongst the tied groups (the strategy we use in our numerical examples
in $\S$\ref{sec:applications}) or else by using a secondary tie-breaking criterion. 

Performance of $k$-NN depends of the choice of $k$: small $k$ allows for classification
boundaries which are flexible
but which have a tendency to overfit, with the opposites true when $k$ is large.
It also depends on the choice of distance measure.  Since we are dealing with compositional data 
we shall use the $\alpha$-metric
(\ref{adist}), denoting such an approach $k$-NN($\alpha$), so $k$-NN(0) indicates the
LRA approach of using $k$-NN with Aitchison's distance (\ref{aitchisondistance}), 
while $k$-NN(1) indicates the EDA approach of using $k$-NN based on Euclidean distance.

We can equally easily use any of many other possible distance measures.  For sake of 
comparing performance with the
$\alpha$-metric we also consider one alternative, namely the following variant of 
the Jensen-Shannon divergence:
\begin{eqnarray} \label{ESOV}
  \text{ESOV}({\bf x},{\bf y})=\sqrt{ \sum_{i=1}^D\left( x_i\log{\frac{2x_i}{x_i+y_i}}+
  y_i\log{\frac{2y_i}{x_i+y_i}} \right) }.
\end{eqnarray}
We use the notation ESOV after \citet{endres2003} and \citet{osterreicher2003} who
independently proved that (\ref{ESOV}) satisfies the triangle inequality and thus is a
metric. As with the $\alpha$-metric (\ref{adist}), the ESOV metric (\ref{ESOV}) is well
defined even when zero values are present. We denote the $k$-NN classifier based on 
metric (\ref{ESOV}) by $k$-NN$_\text{ESOV}$.

\section{Applications of compositional classification}
\label{sec:applications}
We will show four examples of applications of the proposed compositional discrimination
techniques. In all cases we used real data sets, two of them having observations
with zero values in some of the
components, and the other two data sets having no
zero values. The two benchmarks for comparison will be when
$\alpha=0$, which results in LRA, and when $\alpha=1$, which results in EDA.    

We performed RDA$\left(\alpha,\lambda,\gamma \right)$, $k$-NN($\alpha$), 
varying $\alpha$ in steps of 0.05
between -1 and 1 for datasets without zeros and
between 0.05 and 1 for datasets with zeros (since in such circumstances the
$\alpha$-transformation and $\alpha$-metric are not defined for $\alpha \leq 0$), and
varying the values of $\lambda$ and $\gamma$ in steps of 0.1 between 0 and 1. 

To estimate the rate of correct classification in out-of-sample
prediction we used cross validation.  This involves dividing the set of $n$ observations
into training and test sets of size $n_\text{train}$ and $n_\text{test}$ respectively,
training the classifier on the training set, then evaluating its prediction
accuracy of the test set.  In view of the samples having groups with quite variable
numbers of observations we used stratified random sampling to ensure that the training
sets were representative of the test sets, and to arrange 
that all groups were represented in the test
set.  In particular, we randomly divided the samples into training and test sets so that
\begin{eqnarray*}
  \frac{n_i}{n} \approx 
  \frac{n_{i,\text{train}}}{n_{\text{train}}} \approx
  \frac{n_{i,\text{test}}}{n_\text{test}},
\end{eqnarray*}
where $n_i$, $n_{i,\text{train}}$ and $n_{i,\text{test}}$ are the sample sizes of the $i$th group 
in the full, training and test samples, respectively.  We
then estimated the rate of correct classification by
\begin{eqnarray} \label{qu}
  q=\frac{c}{n_\text{test}},
\end{eqnarray}
where $c$ is the number of observations in the test set correctly classified and
$n_\text{test}$ is the test sample size.

For each of the classifiers RDA$\left(\alpha,\lambda,\gamma \right)$ and $k$-NN($\alpha$), 
the steps can be summarised as follows
\begin{enumerate}
  \item[Step 1.] Partition the sample into training and test sets using stratified random
    sampling.
  \item[Step 2.] For each combination of values of the free parameters ($\alpha,\lambda,\gamma$
    for RDA; $\alpha, k$ for $k$-NN($\alpha$); 
    train the classifier on the training set.
  \item[Step 3.] Apply the classifiers to the test set, and calculate $q$ in (\ref{qu}).
  \item[Step 4.] Repeat steps $1-3$ a large number, say $B$, times, then 
    estimate the rate of correct classification as the average of the $q$s in Step 3.
\end{enumerate}

For the calculations in the following section we took $B = 200$ which gave estimates of
the rate of correct classification with small standard errors at reasonable computational
cost.

\subsection{Examples} 
%

We will now introduce four datasets to investigate the performance of the supervised
classification techniques described in \S\ref{sec:techniques}.  The datasets come from
different fields, namely ecology, forensic science, hydrochemistry and economics.

\subsubsection*{Example 1: Fatty acid signature data (contains zero values)}

This is a dataset described in \citep{stewart2011} (itself an updated version of a
dataset from \citep{iverson2004}) which contains observations of $n=2110$ fish of $g=28$
different species, each observation being a composition with $D=40$ components that
characterises the fatty acid signature of the fish.  A special feature of this dataset
is that it contains many zero values (3506 components, across all observations, are
zero) which rules out use of the log-ratio transformation (\ref{aitchisonclr}).  Table
\ref{diet} shows the number of observations in each group, and the number of observations for
which at least one component is zero.  Table \ref{zeros1} shows the proportion of
observations which have zeros in each of the components.  For this example, for the
cross validation we used a test set of $n_\text{test}=165$ observations ($7.8\%$ of the
full sample).

\begin{table}[h]
\begin{small}
\begin{center}
\begin{tabular}{l|c|l|c|l|c} \hline
Species           &  Sample size  &  Species     & Sample size  & Species          & Sample size  \\  \hline  \hline
Butterfish        &  35(30)       &  Mackerel    &  34(23)      & Snake Blenny     &  18(12)    \\
Capelin           &  165(145)     &  Ocean Pout  &  31(31)      & Squid            &  18(17)    \\ 
Cod               &  147(121)     &  Plaice      &  148(120)    & Thorny Skate     &  74(74)    \\ 
Gaspereau         &  70(69)       &  Pollock     &  57(49)      & Turbot           &  20(20)    \\
Haddock           &  148(134)     &  Red Hake    &  25(24)      & White Hake       &  75(71)    \\
Halibut           &  13(11)       &  Redfish     &  84(74)      & White Flounder   &  90(80)    \\
Herring           &  247(231)     &  Sandlance   &  124(101)    & Winter Skate     &  40(39)    \\
Lobster           &  21(21)       &  Shrimp      &  122(87)     & Witch Flounder   &  24(24)    \\
Longhorn Sculpin  &  70(69)       &  Silver Hake &  70(58)      & Yellow Tail      &  118(103)  \\
Lumpfish          &  22(13)       &              &              &                  &            \\  \hline
\end{tabular}       
\caption{Sample sizes of the different species of the fatty acid data. The number inside the parentheses shows how many observations have at least one zero element.}
\label{diet}
\end{center}
\end{small} 
\end{table}
\begin{table}[h]
\begin{small}
\begin{center}
\begin{tabular}{c|c|c|c|c|c|c|c|c|c|c} \hline 
Component & 1st & 2nd & 3rd & 4th & 5th & 6th & 7th & 8th & 9th & 10th \\ \hline 
Percentage of zeros & 0.00\% & 0.00\% & 0.00\% & 6.54\% & 0.28\% & 9.86\% & 9.10\% & 4.88\% & 65.36\% & 2.94\% \\
\hline \hline
Component & 11th & 12th & 13th & 14th & 15th & 16th & 17th & 18th & 19th & 20th \\ \hline
Percentage of zeros & 0.00\% & 0.00\% & 0.00\% & 0.00\% & 6.78\% & 2.32\% & 0.62\% & 0.00\% & 3.51\% & 0.05\% \\
\hline \hline
Component & 21st & 22nd & 23rd & 24th & 25th & 26th & 27th & 28th & 29th & 30th \\ \hline 
Percentage of zeros & 2.65\% & 0.09\% & 0.00\% & 0.05\% & 1.80\% & 1.66\% & 0.00\% & 0.33\% & 0.05\% & 0.00\% \\
\hline \hline
Component & 31st & 32nd & 33rd & 34th & 35th & 36th & 37th & 38th & 39th & 40th \\ \hline 
Percentage of zeros & 0.33\% & 0.5\% & 0.00\% & 27.35\% & 0.00\% & 10.66\% & 0.00\% & 8.91\% & 0.00\% & 0.00\% \\
\hline 
\end{tabular}       
\caption{Fatty acid data: the percentage of observations for which each component is zero.}
\label{zeros1}
\end{center}
\end{small} 
\end{table}


\subsubsection*{Example 2: Forensic glass data (contains zero values)}

In the second example we use the forensic glass dataset \citep{UCI} which has $n=214$
observations from $g=6$ different categories of glass, where each observation is a
composition with $D=8$ components.  The categories which occur are: containers ($13$
observations, 12 of which have at least one zero element), vehicle headlamps ($29$
observations, all with at least one zero value), tableware ($9$ observations, all with at
least one zero value), vehicle window glass ($17$ observations, 16 with at least one zero
value), window float glass ($70$ observations, 69 with at least one zero value) and window
non-float glass ($76$ observations, 72 with at least one zero value). Once again the zeros
rule out the use of LRA.  In total there are $392$ zero values; Table \ref{zeros2} shows
in which components these zeros arise and Table \ref{tab1b} summarises the distribution of
zeros across the observations.  For the cross validation we used a test set consisted of
$n_\text{test} = 30$ observations ($14\%$ of the total sample).
\begin{table}[h]
\begin{center}
\begin{tabular}{c|c|c|c|c} \hline
Components           & Sodium    & Magnesium & Aluminium & Silicon  \\ \hline \hline   
Percentage of zeros  & 0.00\%    & 19.63\%   & 0.00\%    & 0.00\%   \\ \hline 
Components           & Potassium & Calcium   & Barium    & Iron       \\ \hline \hline
Percentage of zeros  & 14.02\%   & 0.00\%    & 82.24\%   & 67.29\%    \\ \hline 
\end{tabular}       
\caption{Forensic glass data: the percentage of observations for which each component is zero.}
\label{zeros2}
\end{center}
\end{table}

\subsubsection*{Example 3: Hydrochemical data (contains no zero values)}
The hydrochemical data set \citep{otero2005} contains compositional observations on
$D=14$ chemicals (H, Na, K, Ca, Mg, Sr, Ba, NH4, Cl, HCO3, NO3, SO4, PO4, TOC) in water
samples from tributaries of the Llobregat river in north-east Spain.  The $n=485$ observations
are in $g=4$ groups according to which tributary they were measured in: 
Anoia ($143$ observations), Cardener ($95$ observations),
Upper Llobregat ($135$ observations) or Lower Llobregat ($112$ observations).  For the
cross validation in this example we used a training set of size
$n_\text{test} = 165$ ($34\%$
of the total sample size).

\subsubsection*{Example 4: National income data (contains no zero
values)} This final example is an economics data set \citep{larrosa2003} containing
compositional observations for $n=56$
countries with $D=5$ components reflecting the proportion of
capital allocated in production assets, residential buildings, non-residential
buildings, other buildings, and transportation equipment.
The countries are categorised into $g=5$ groups according to income levels
and membership of the Organization for Economic Co-operation and Development (OECD); the
groups are ``low income'' ($10$ countries), ``lower middle income'' ($12$ countries),
``upper middle income'' ($9$ countries), ``high income and OECD member'' ($21$ countries),
and ``high income and non-OECD member'' ($4$ countries).  For the cross validation, we
used a test set of
$n_\text{test} = 10$ observations ($17.9\%$ of the total sample).   

\subsection{Results}
\begin{figure}[!ht]
\centering
\begin{tabular}{ccc}
\multicolumn{3}{c}{Example 1 (Fatty acid signature data)} \\ 
\includegraphics[scale=0.32,trim=0 20 0 20]{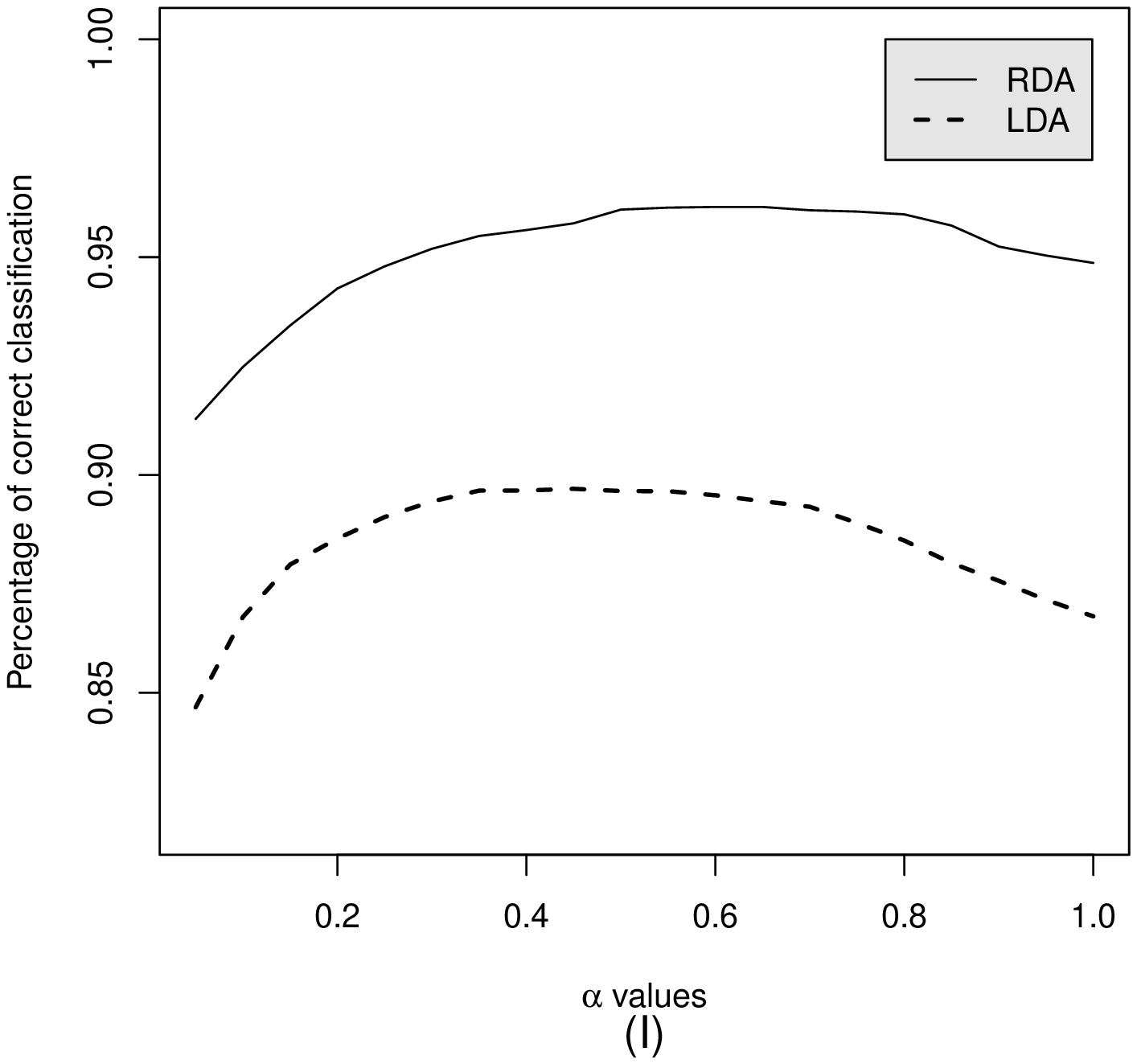} &
\includegraphics[scale=0.32,trim=0 20 0 20]{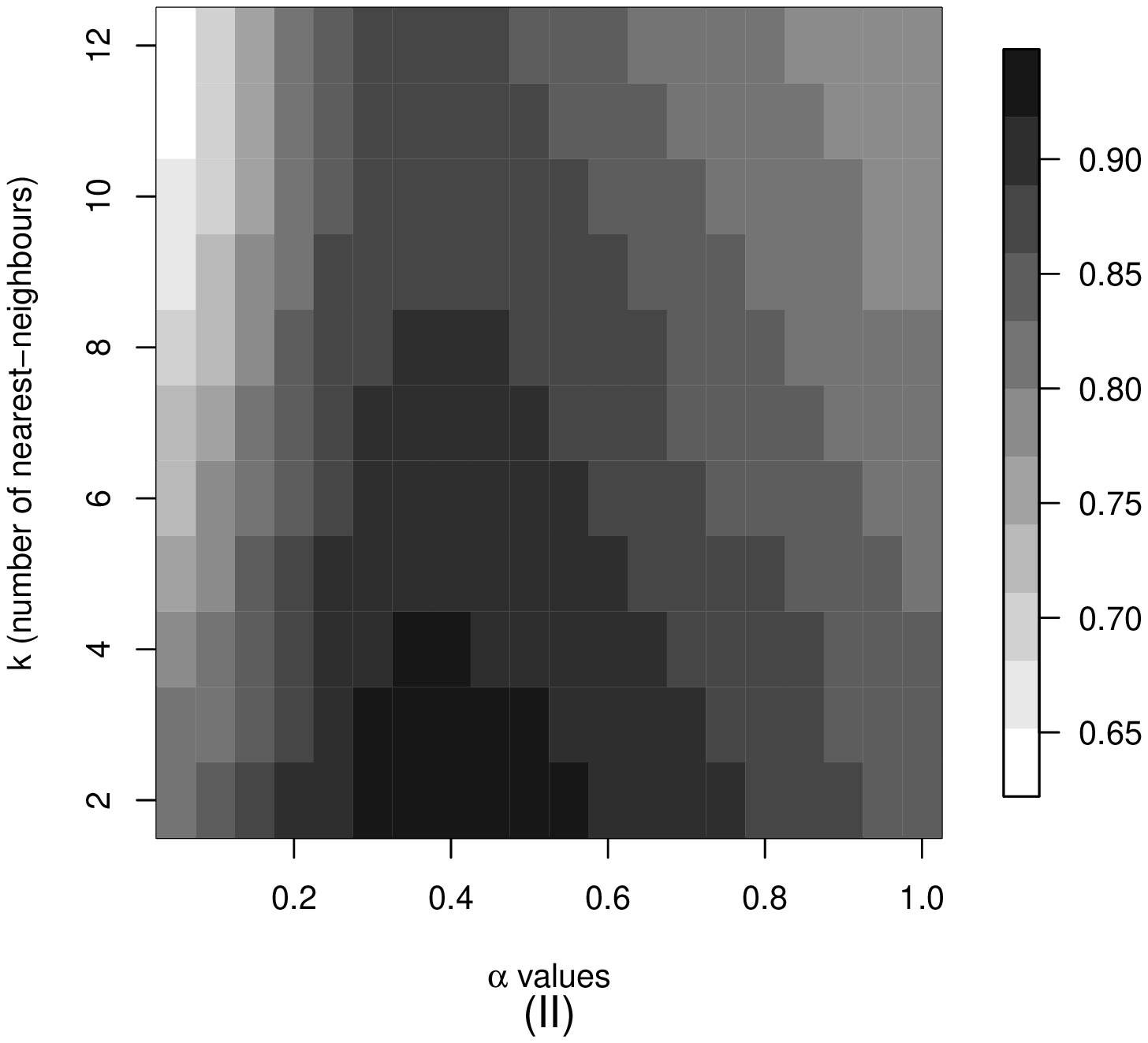} &
\includegraphics[scale=0.32,trim=0 20 0 20]{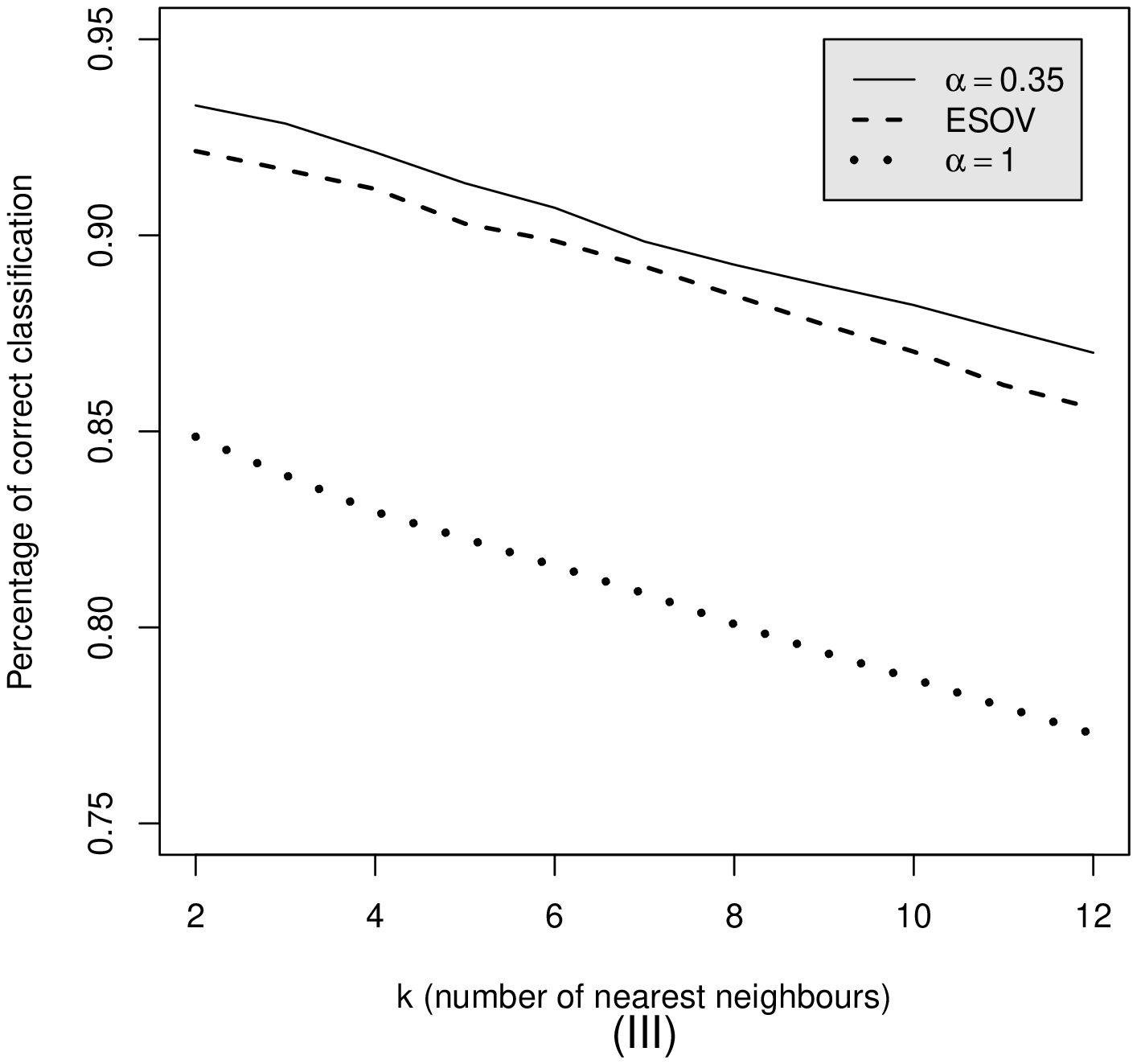} \\
\multicolumn{3}{c}{Example 2 (Forensic glass data)} \\ 
\includegraphics[scale=0.32,trim=0 20 0 20]{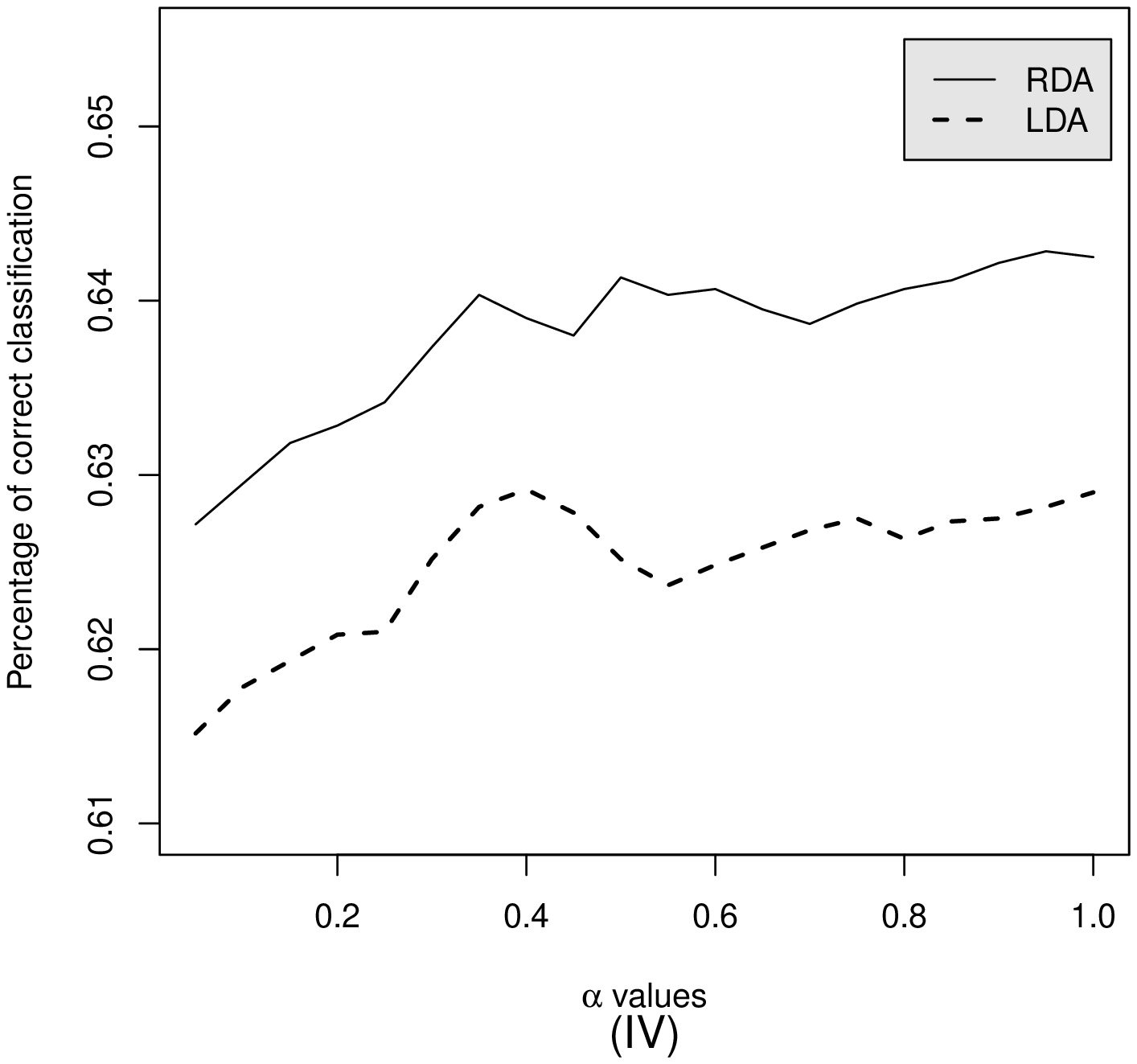} &
\includegraphics[scale=0.32,trim=0 20 0 20]{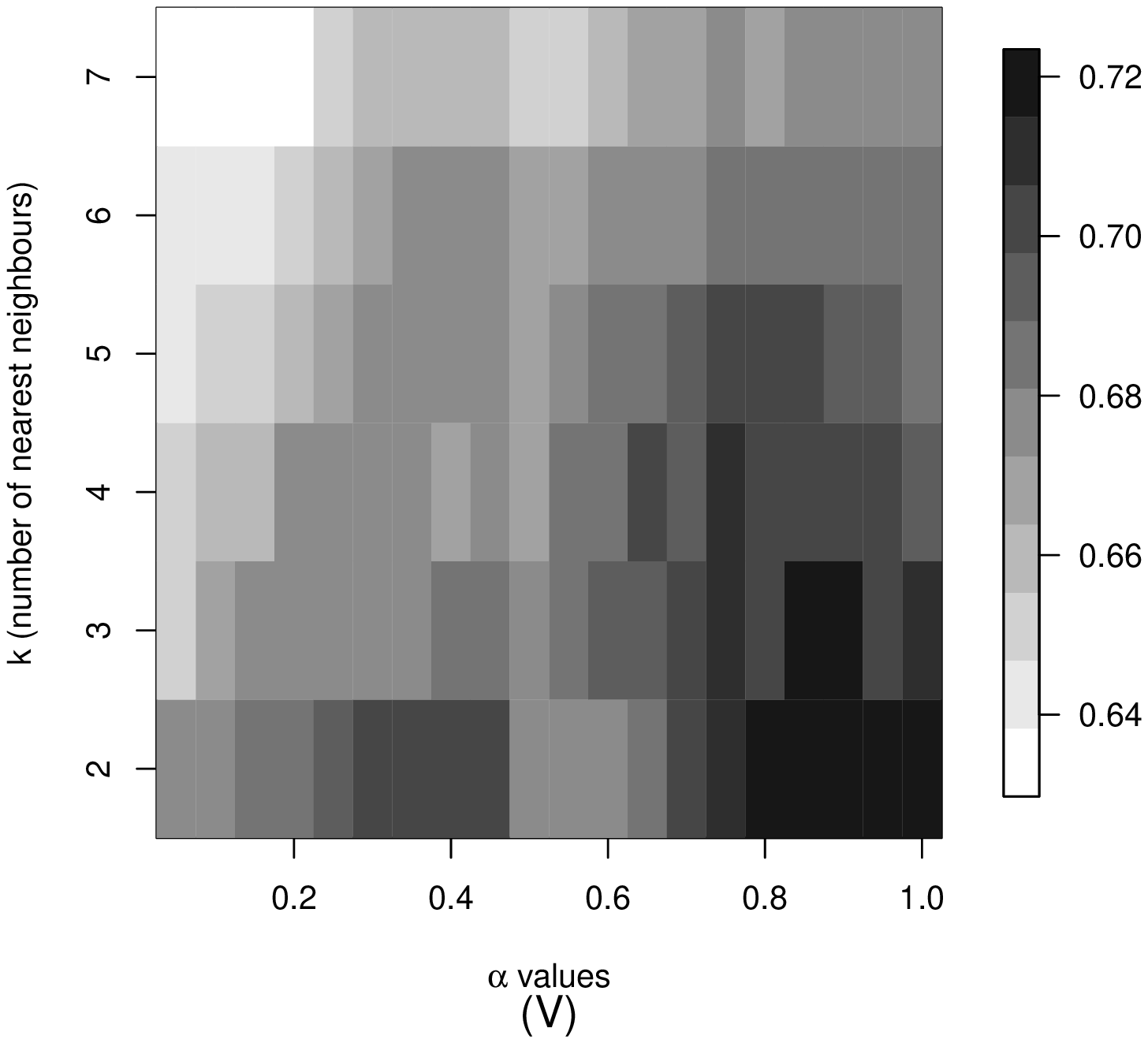}  &
\includegraphics[scale=0.32,trim=0 20 0 20]{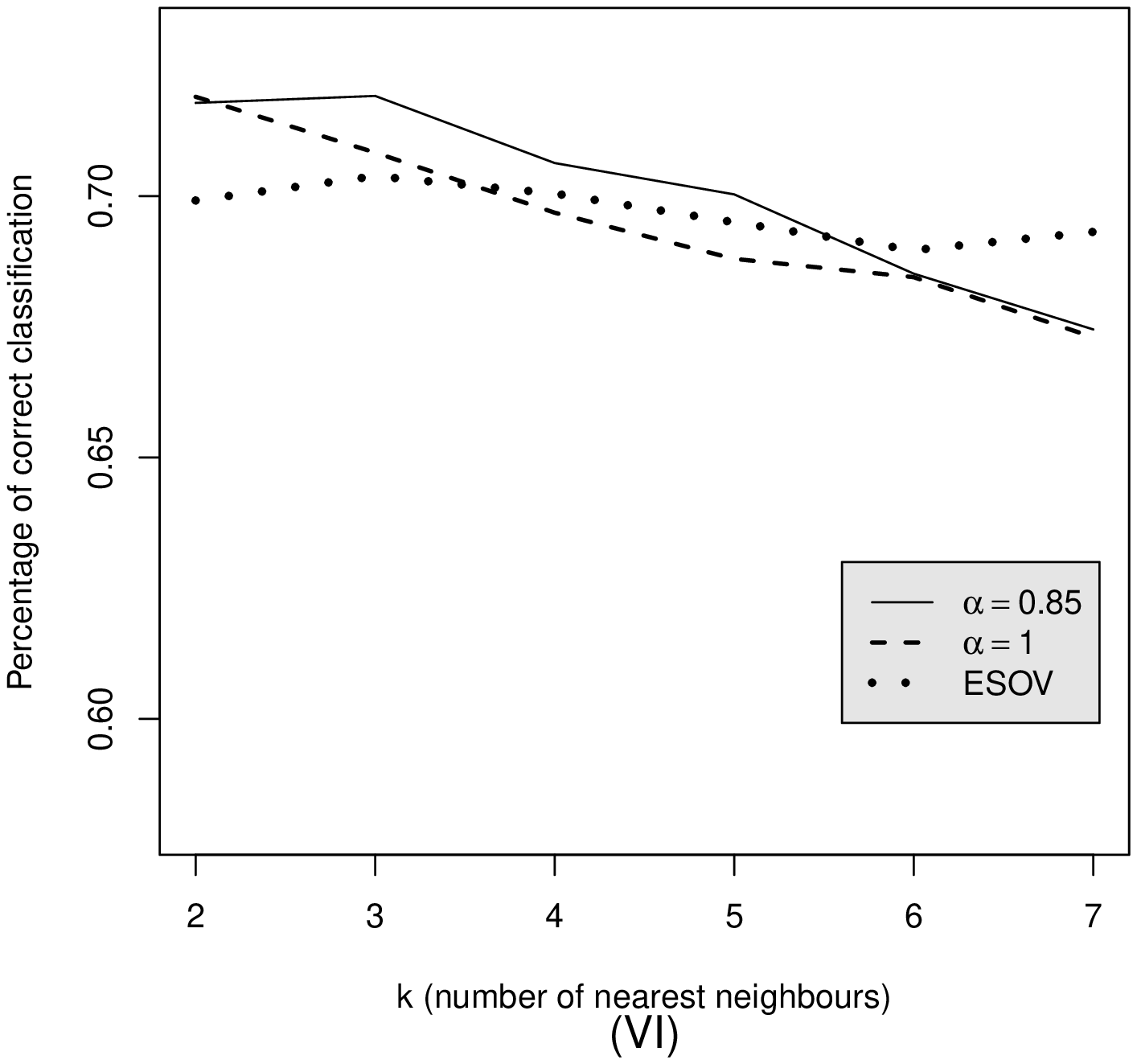} \\
\multicolumn{3}{c}{Example 3 (Hydrochemical data)} \\ 
\includegraphics[scale=0.32,trim=0 20 0 20]{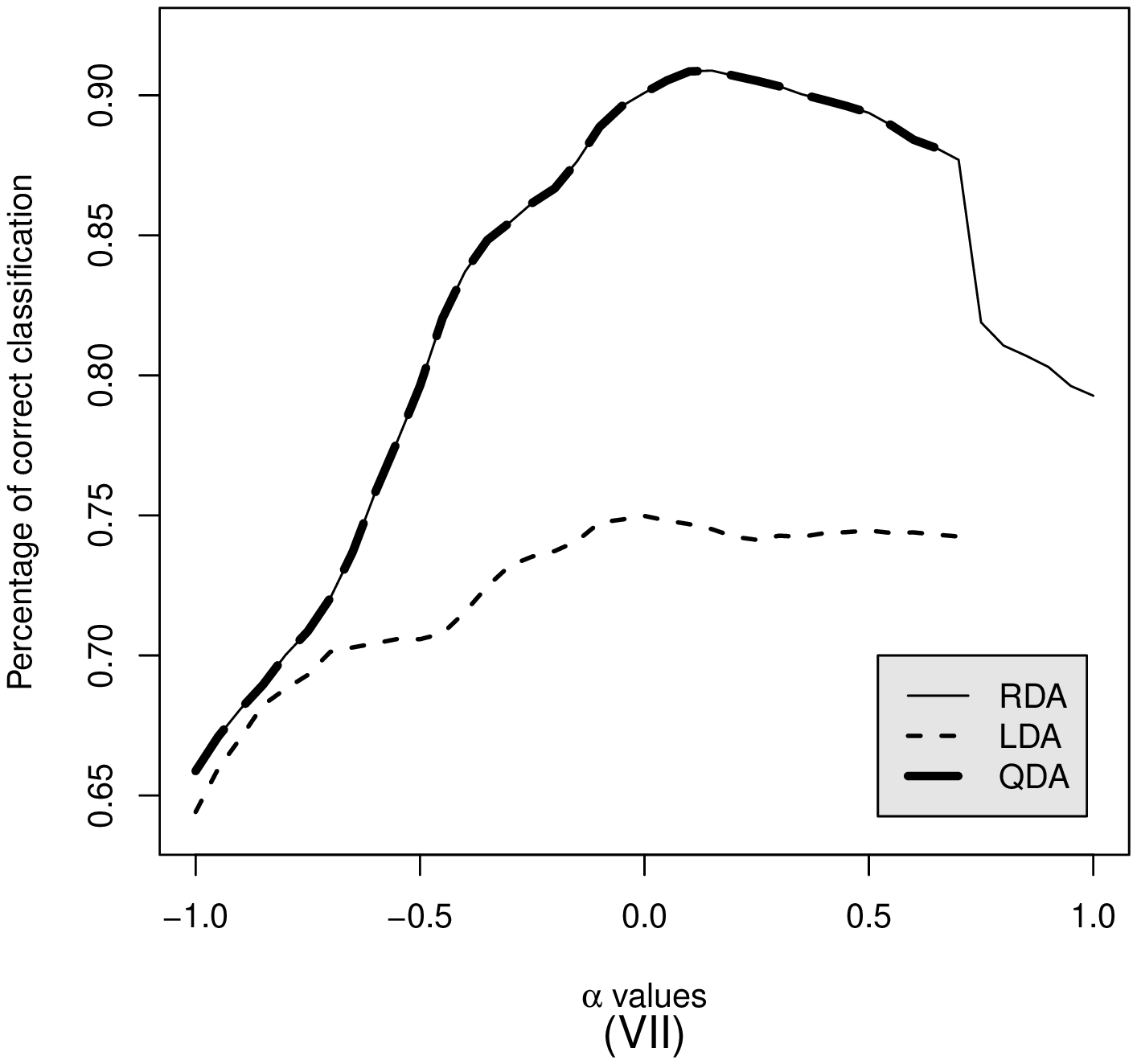}   &
\includegraphics[scale=0.32,trim=0 20 0 20]{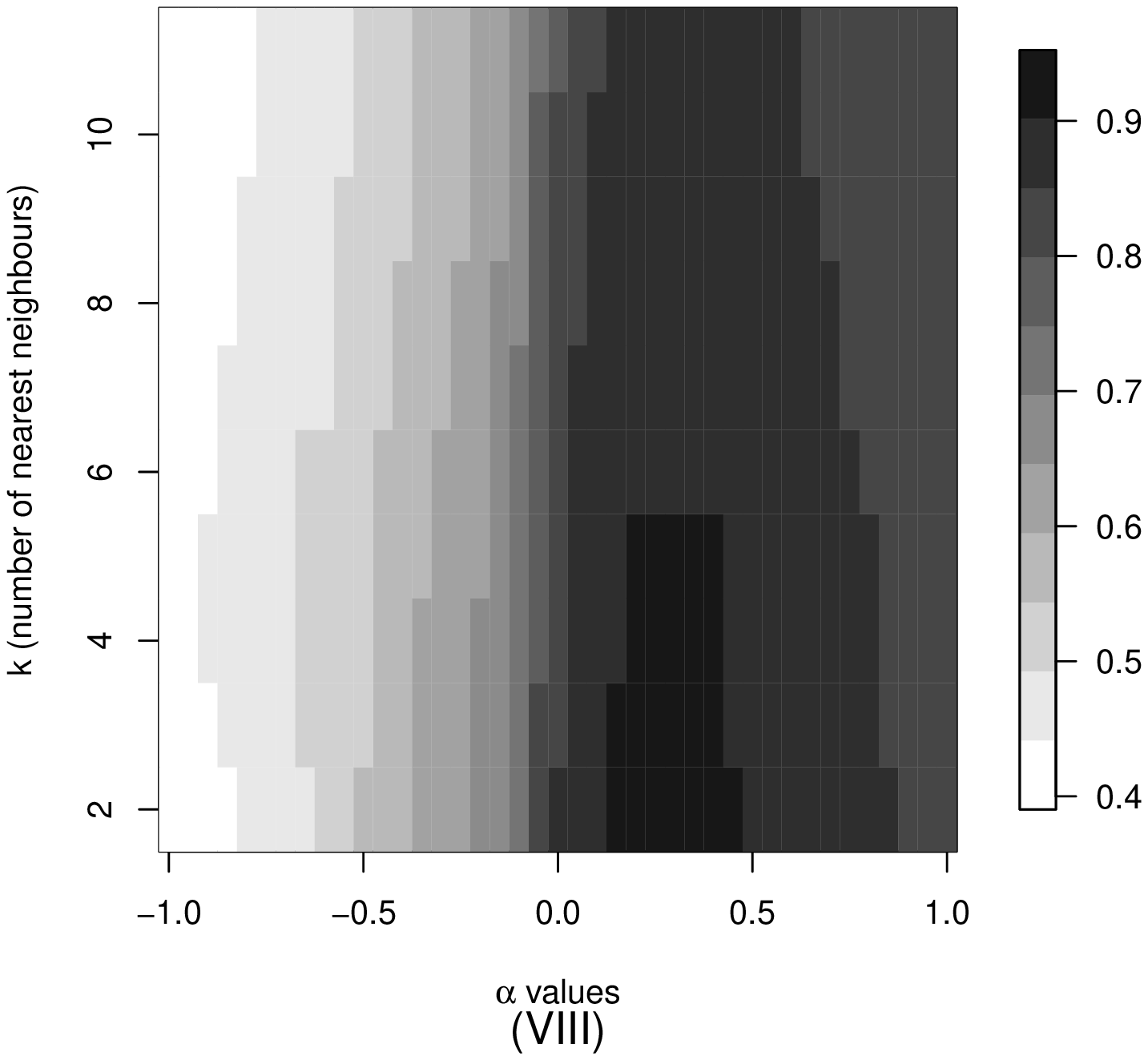} & 
\includegraphics[scale=0.32,trim=0 20 0 20]{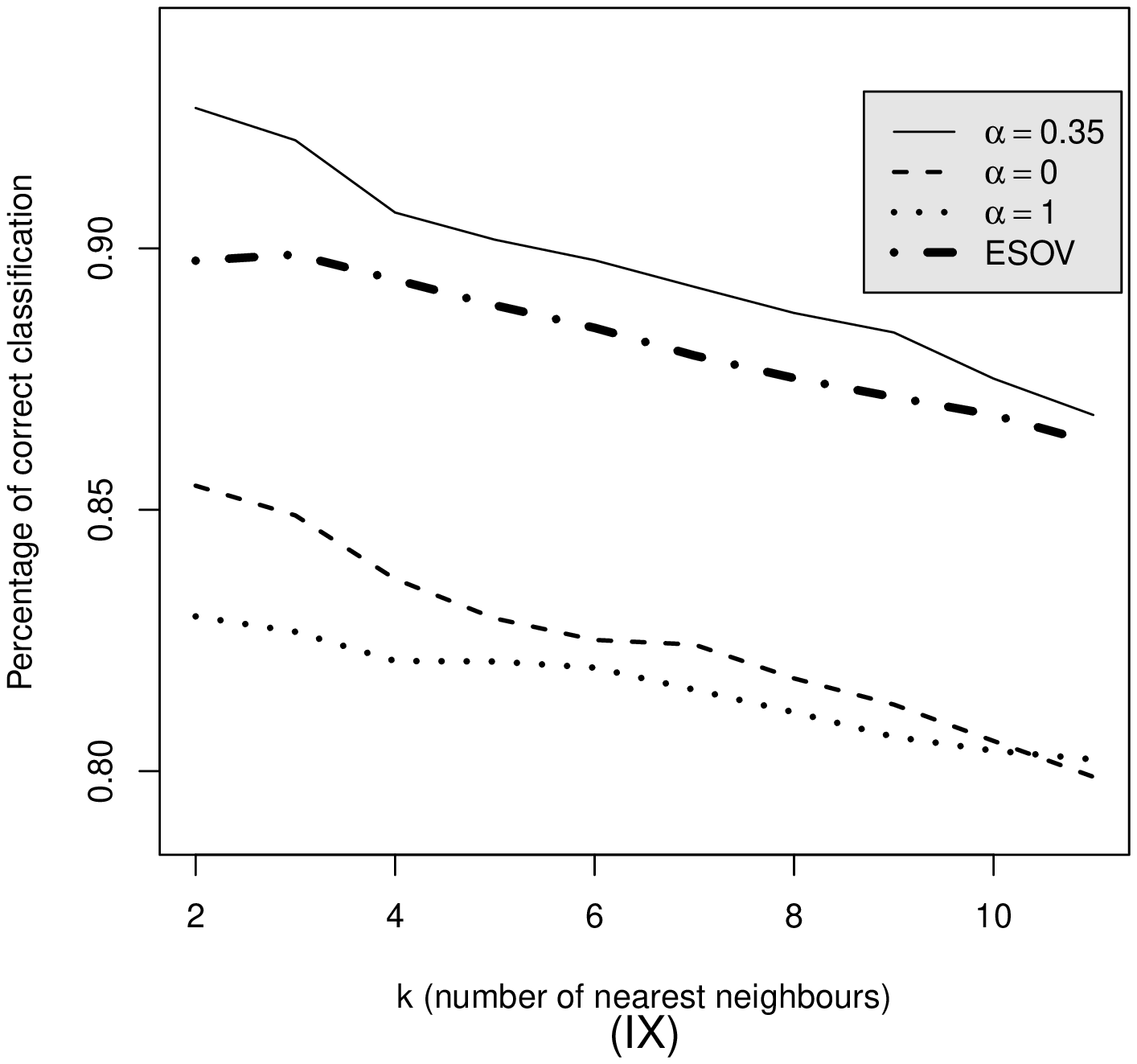} \\
\multicolumn{3}{c}{Example 4 (National income data)} \\ 
\includegraphics[scale=0.32,trim=0 20 0 20]{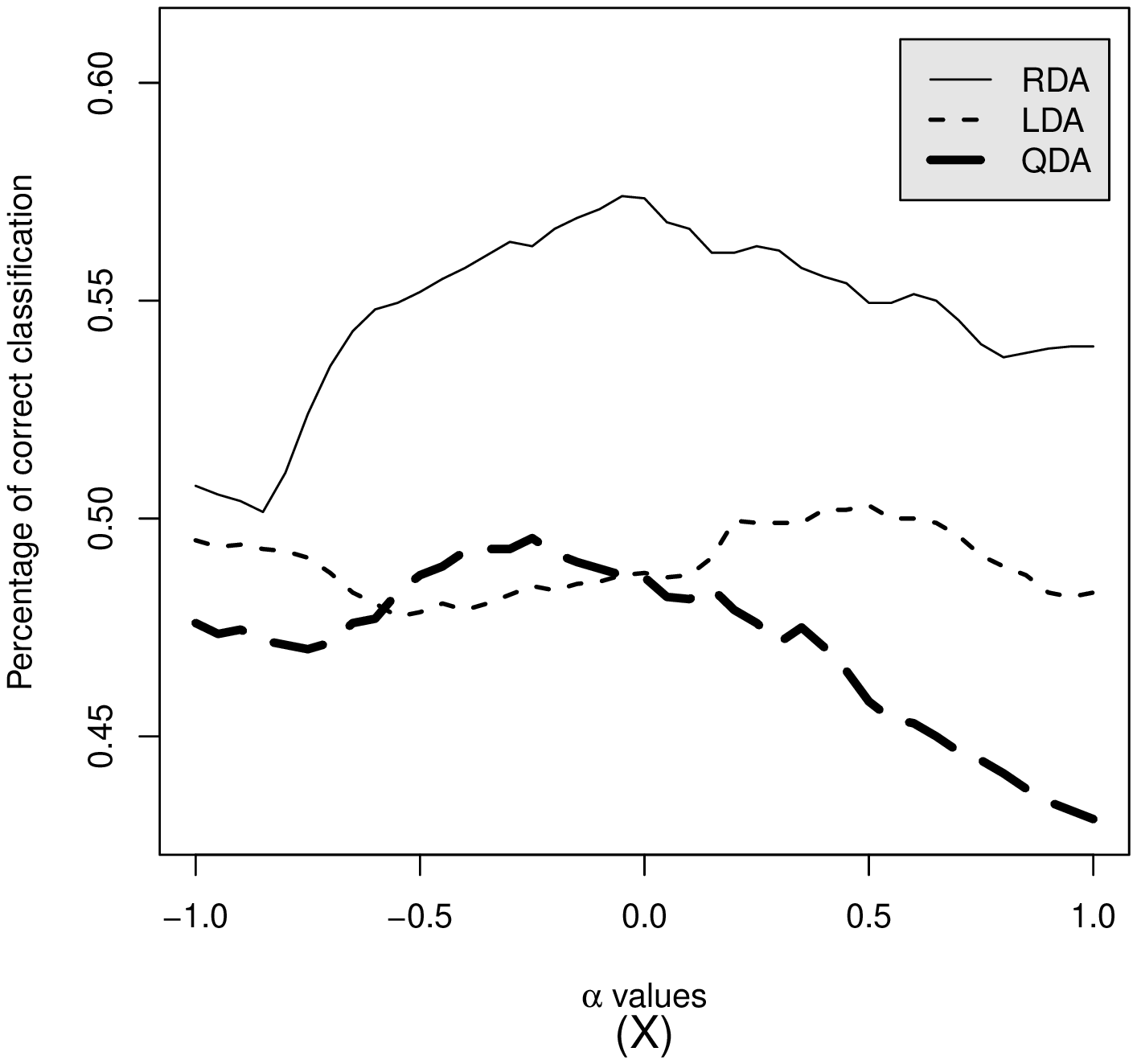}   &
\includegraphics[scale=0.32,trim=0 20 0 20]{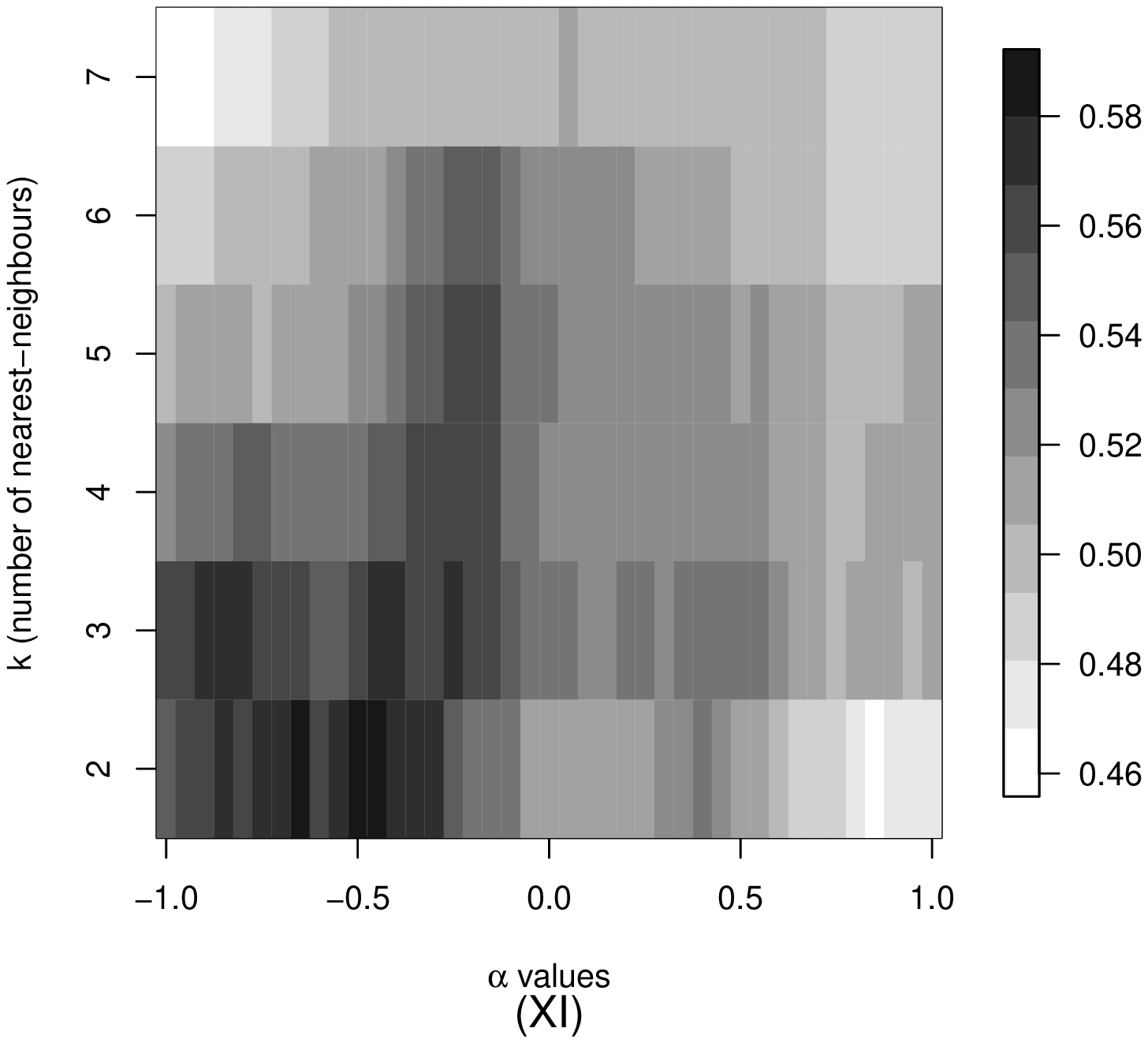}   &
\includegraphics[scale=0.32,trim=0 20 0 20]{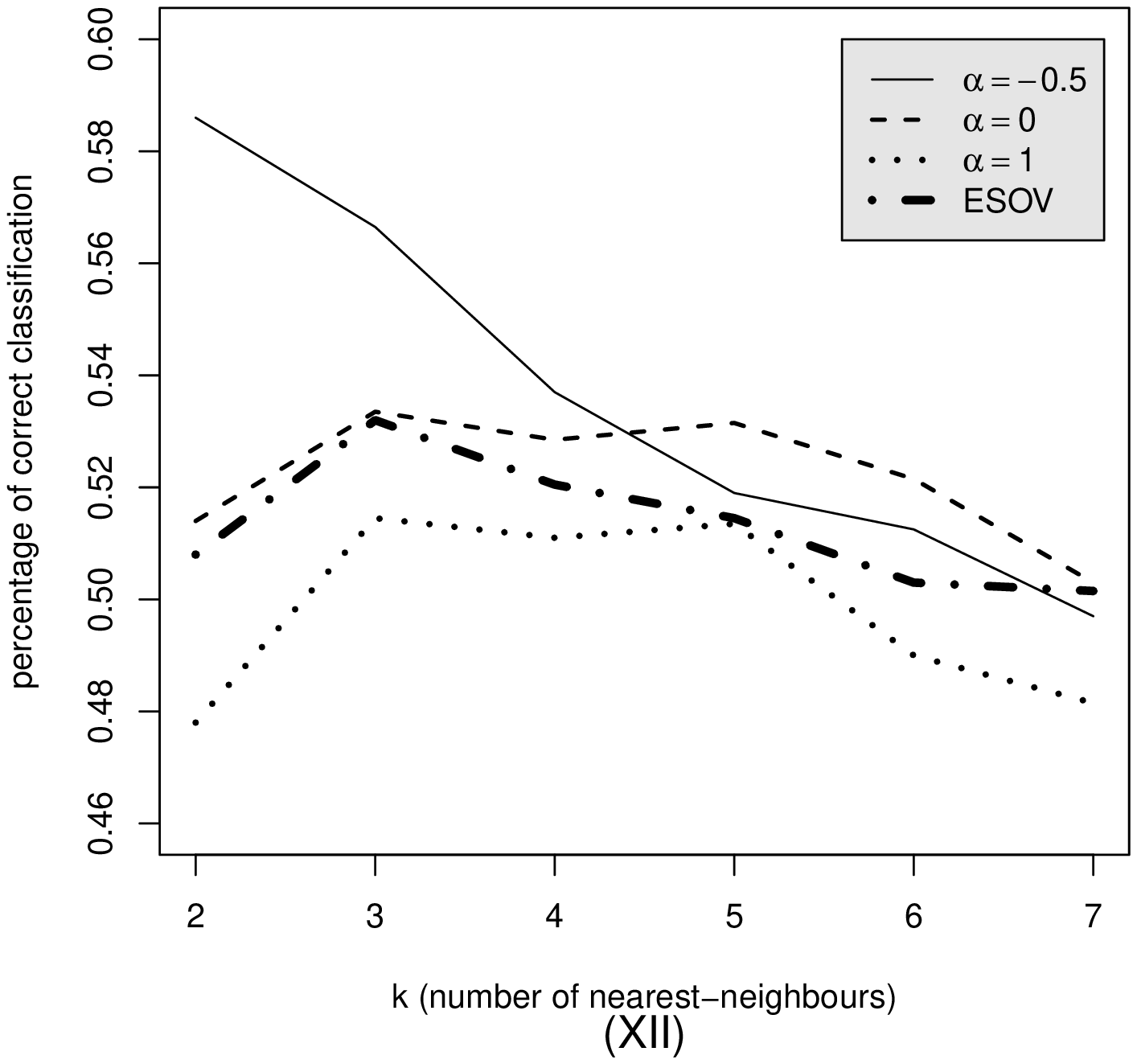} 
\end{tabular}
\caption{All graphs contain the estimated rate of correct classification for the different methods. The first column refers to LDA, QDA and RDA as a function of $\alpha$. The second column contains the heat plots of the $k$-NN algorithm as a function of $\alpha$ and $k$, the nearest neighbours. The graphs in the third column present the results of the $k$-NN algorithm of the $\alpha$-metric for some specific values of $\alpha$.}
\label{figure}
\end{figure}

\begin{figure}[ht]
\centering
\begin{tabular}{ccc}
  \multicolumn{3}{l}{}      \\
\includegraphics[scale=0.3,trim=0 20 0 20]{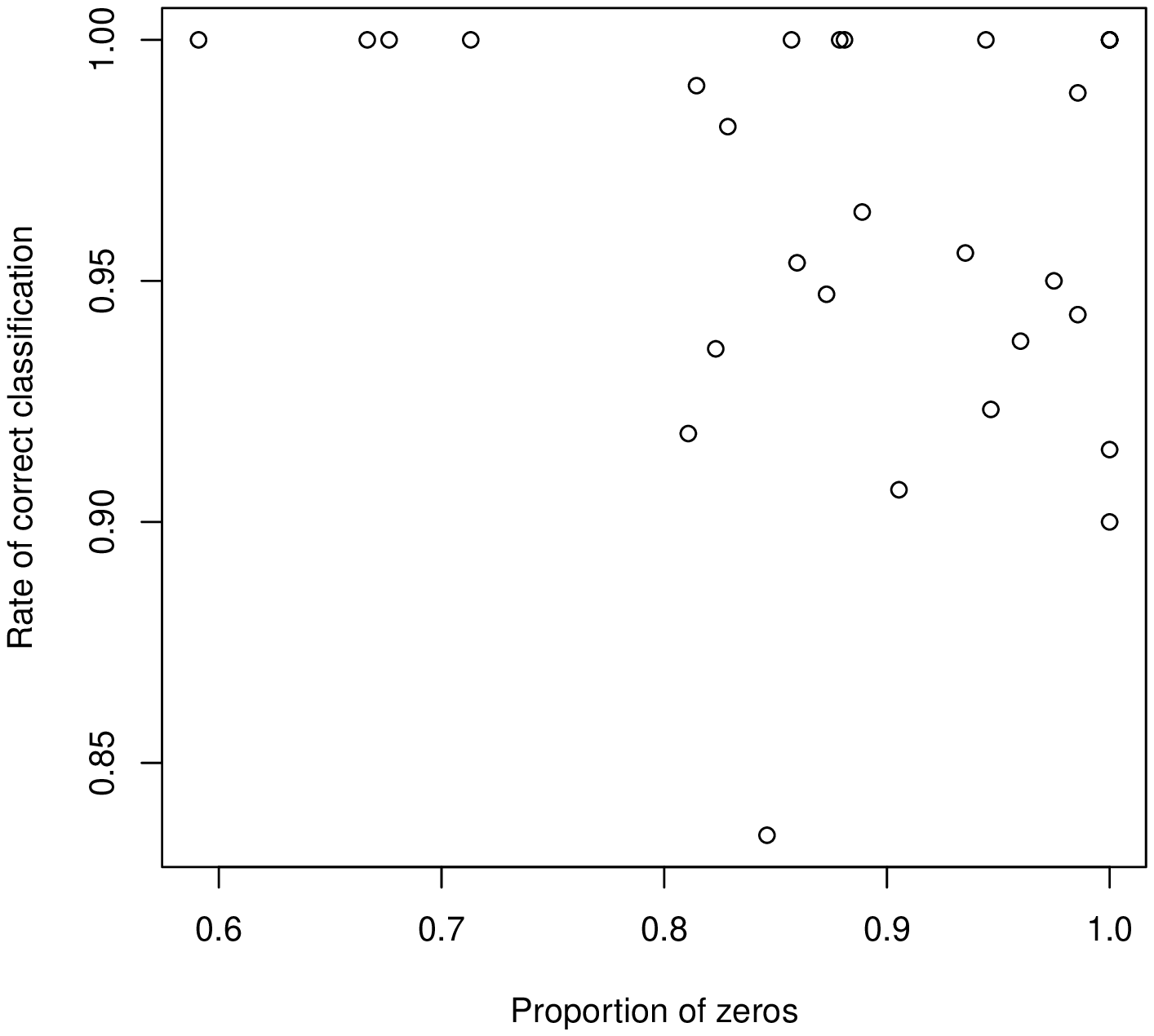} &
\includegraphics[scale=0.3,trim=0 20 0 20]{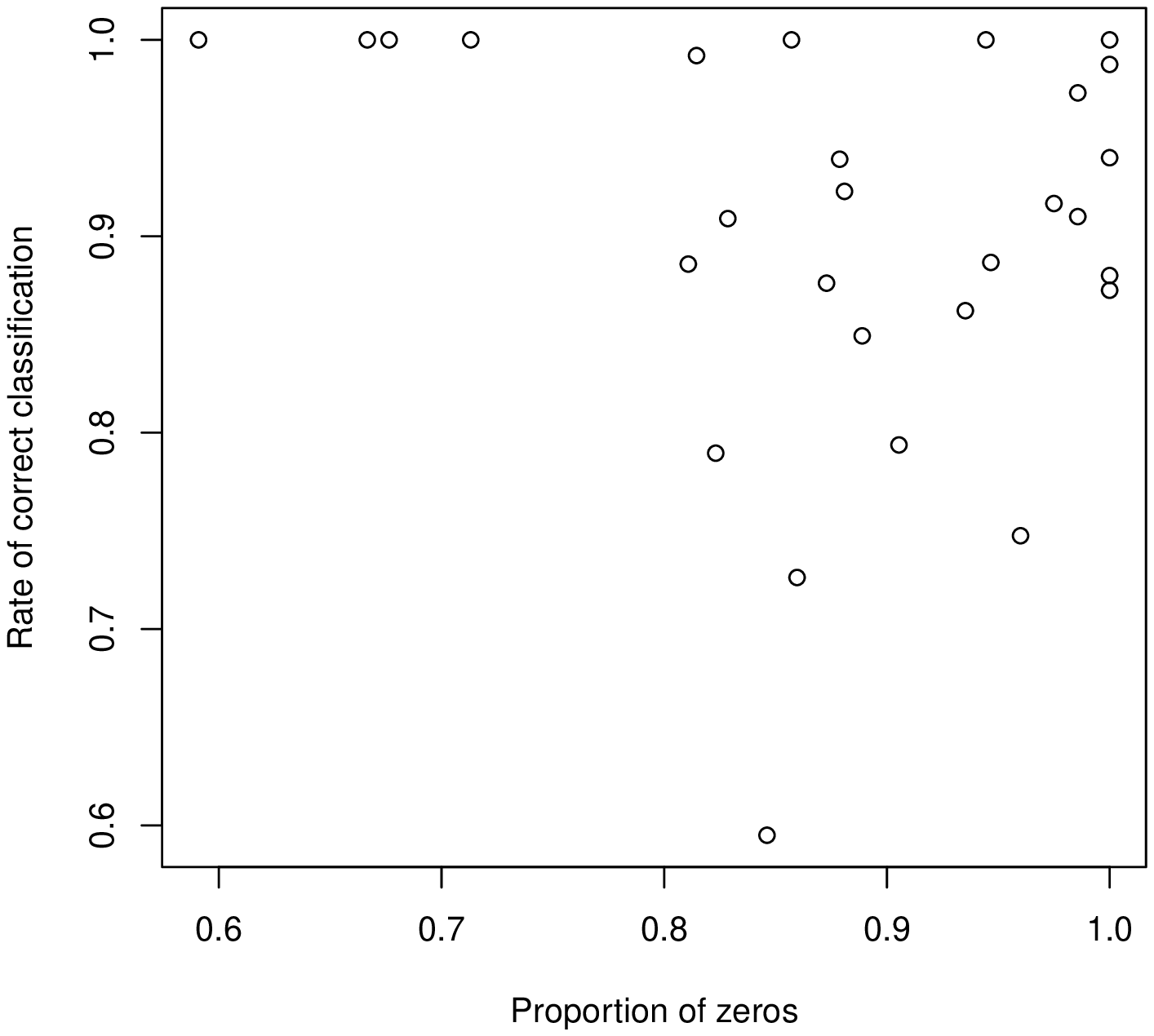} &
\includegraphics[scale=0.3,trim=0 20 0 20]{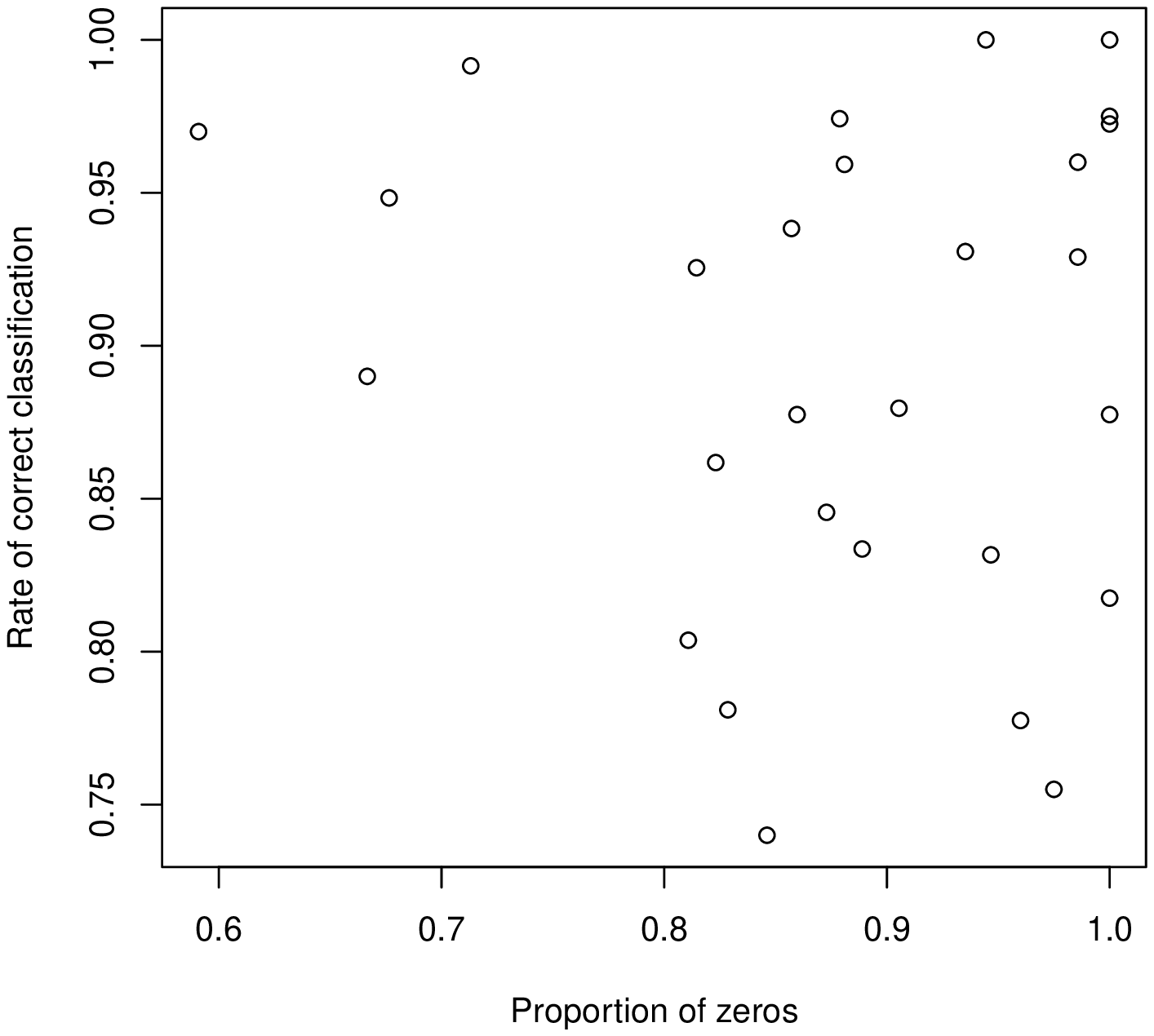} \\
RDA$\left(0.6,0.9,0.7\right)$  & LDA$\left(0.45\right)$  & $2$-NN$\left(0.35\right)$ 
\end{tabular}
\caption{Fatty acid signature data: the estimated rate of correct 
classification accuracy by group versus the the proportion of observations within the group that contain at least one zero.}
\label{figure2}
\end{figure}

This section contains results from applying the methods of \S\ref{sec:techniques} to the
four compositional datasets described above.  Results are summarised in Figures
\ref{figure} and \ref{figure2} and Tables \ref{tab1}-\ref{tab2}.  The Tables show results for
$\alpha=1$, $\alpha=0$, and for $\alpha$ free in [-1,1], in each case for the values of
free parameters that maximise the estimated rate of correct classification.

\begin{table}[!ht]
\begin{center}
\begin{tabular}{c|c|c|c} 
\multicolumn{4}{c}{\textbf{Example 1 (Fatty acid signature data)}} \\ \hline
                              &  Estimated rate of       &                           & Estimated rate of       \\
Method                        &  correct classification  & Method                    & correct classification  \\ \hline \hline
RDA$\left(0.6,0.9,0.7\right)$ &  0.962(0.014)            & RDA$\left(1,0.8,1\right)$ & 0.949(0.016)            \\   
LDA$\left(0.45\right)$        &  0.897(0.022)            & LDA$\left(1\right)$       & 0.868(0.024)            \\ \hline
$2$-NN$\left(0.35\right)$     &  0.933(0.020)            & $2$-NN$\left(1\right)$    & 0.849(0.027)            \\ \hline
$2$-NN$_{ESOV}$               &  0.921(0.019)            &                           &                         \\ \hline \hline 
\multicolumn{4}{c}{\textbf{Example 2 (Forensic glass data)}} \\ \hline
                             &  Estimated rate of       &                           & Estimated rate of        \\
Method                       &  correct classification  & Method                    & correct classification   \\ \hline \hline    
RDA$\left(0.95,0.1,1\right)$ &  0.643(0.034)            & RDA$\left(1,0.1,1\right)$ & 0.643(0.034)             \\      
LDA$\left(0.4\right)$        &  0.629(0.034)            & LDA$\left(1\right)$       & 0.629(0.034)             \\ \hline
$3$-NN$\left(0.85\right)$    &  0.719(0.033)            & $2$-NN$\left(1\right)$    & 0.719(0.033)             \\ \hline
$3$-NN$_{ESOV}$              &  0.693(0.033)            &                           &                          \\  \hline \hline  
\end{tabular}
\caption{Estimated rate of correct classification of the different approaches. The standard error appears inside the parentheses.}
\label{tab1}
\end{center}
\end{table}

\begin{table}[!ht]
\begin{center}
\begin{tabular}{c|c|c|c|c|c}  \hline
                              & \multicolumn{5}{c}{Number of zeros}    \\ \hline 
Method                        & 0 (12.27\%)     & 1 (45.69\%)    & 2 (2.61\%)    & 3 (9.38\%)     & 4-8 (10.5\%)    \\  \hline \hline
RDA$\left(0.6,0.90.71\right)$ & 0.956(0.043) & 0.963(0.021) & 0.963(0.030) & 0.976(0.040) & 0.949(0.053)  \\
RDA$\left(1,0.8,1\right)$     & 0.951(0.045) & 0.962(0.022) & 0.941(0.037) & 0.942(0.063) & 0.911(0.066)  \\ \hline
LDA$\left(0.45\right)$        & 0.875(0.065) & 0.925(0.029) & 0.881(0.051) & 0.872(0.091) & 0.855(0.093)  \\
LDA$\left(1\right)$           & 0.882(0.060) & 0.898(0.034) & 0.842(0.053) & 0.822(0.099) & 0.813(0.096)  \\ \hline
$2$-NN$\left(0.35\right)$     & 0.923(0.054) & 0.938(0.025) & 0.923(0.039) & 0.963(0.047) & 0.922(0.064)  \\
$2$-NN$\left(1\right)$        & 0.844(0.075) & 0.853(0.036) & 0.853(0.058) & 0.874(0.082) & 0.803(0.100)  \\ \hline
$2$-NN$_{ESOV}$               & 0.918(0.062) & 0.928(0.030) & 0.913(0.046) & 0.962(0.050) & 0.880(0.086)  \\ \hline
\end{tabular}
\caption{Fatty acid signature data: classification accuracy by number of zeros.  The
estimated rate of correct classification is shown (with standard errors in parentheses).}
\label{tab2a}
\end{center}
\end{table}

\begin{table}[!ht]
\begin{center}
\begin{tabular}{c|c|c|c|c|c}  \hline
                             & \multicolumn{5}{c}{Number of zeros}     \\ \hline 
Method                       &  0 (3.27\%)   & 1 (29.44\%)  & 2 (50.93\%)  & 3 (13.55\%)  & 4 (2.80\%)    \\  \hline \hline
RDA$\left(0.95,0.1,1\right)$ &  0.421(0.433) & 0.582(0.173) & 0.668(0.111) & 0.787(0.233) & 0.402(0.463)  \\
RDA$\left(1,0.1,1\right)$    &  0.428(0.435) & 0.585(0.174) & 0.665(0.112) & 0.788(0.233) & 0.397(0.459)  \\ \hline
LDA$\left(0.4\right)$        &  0.404(0.431) & 0.536(0.165) & 0.636(0.120) & 0.869(0.194) & 0.689(0.412)  \\
LDA$\left(1\right)$          &  0.397(0.430) & 0.523(0.162) & 0.673(0.110) & 0.790(0.230) & 0.463(0.462)  \\ \hline
$3$-NN$\left(0.85\right)$    &  0.307(0.394) & 0.713(0.160) & 0.717(0.108) & 0.925(0.146) & 0.387(0.429)  \\
$2$-NN$\left(1\right)$       &  0.568(0.441) & 0.715(0.160) & 0.712(0.114) & 0.870(0.178) & 0.387(0.429)  \\ \hline
$3$-NN$_{ESOV}$              &  0.477(0.447) & 0.644(0.164) & 0.764(0.097) & 0.731(0.243) & 0.387(0.429)  \\ \hline
\end{tabular}
\caption{Forensic glass data: classification accuracy by number of zeros.  The
estimated rate of correct classification is shown (with standard errors in parentheses).}
\label{tab1b}
\end{center}
\end{table}

\begin{table}[!ht]
\begin{center}
\begin{tabular}{c|c|c|c|c|c}
\multicolumn{6}{c}{\textbf{Example 3 (Hydrochemical data)}} \\ \hline
                           &  Estimated       &             & Estimated       &        & Estimated       \\
Method                     &  rate of         & Method      & rate of         & Method & rate of         \\  
                           &  correct         &             & correct         &        & correct         \\
                           &  classification  &             & classification  &        & classification  \\ \hline \hline 
RDA$\left(0.15,1,0\right)$ &  0.909(0.02)     & RDA$\left(0,1,0\right)$  & 0.901(0.021)   & RDA$\left(1,0.9,0.9\right)$ & 0.793(0.029)  \\  
QDA$\left(0.15\right)$     &  0.909(0.02)     & QDA$\left(0\right)$      & 0.901(0.021)   & QDA$\left(1\right)$      &   -          \\
LDA$\left(0\right)$        &  0.750(0.031)    & LDA$\left(0\right)$      & 0.750(0.031)   & LDA$\left(1 \right)$     &   -          \\ \hline
$2$-NN$\left(0.25\right)$  &  0.927(0.020)    & $2$-NN$\left(0\right)$   & 0.855(0.026)   & $2$-NN$\left(1\right)$   & 0.830(0.027)  \\ \hline
$3$-NN$_{ESOV}$            &  0.899(0.021)    &                          &                &                          &        \\ \hline \hline
\multicolumn{6}{c}{\textbf{Example 4 (National income data)}} \\ \hline
                              &  Estimated      &           & Estimated      &         & Estimated      \\
Method                        &  rate of  & Method    & rate of  & Method  & rate of  \\  
                              &  correct        &           & correct        &         & correct        \\
                              &  classification &           & classification &         & classification \\ \hline \hline 
RDA$\left(-0.05,0.5,0\right)$ &  0.574(0.035)  & RDA$\left(0,0.5,0\right)$  & 0.574(0.035)  & RDA$\left(1,0.2,0\right)$ & 0.540(0.035) \\  
QDA$\left(-0.25\right)$       &  0.496(0.035)  & QDA$\left(0\right)$      & 0.487(0.035)  & QDA$\left(1\right)$       & 0.431(0.035)   \\
LDA$\left(0.5\right)$         &  0.503(0.035)  & LDA$\left(0\right)$      & 0.488(0.035)  & LDA$\left(1 \right)$      & 0.483(0.035)  \\ \hline
$2$-NN$\left(-0.5\right)$     &  0.586(0.035)  & $3$-NN$\left(0\right)$   & 0.533(0.035)  & $3$-NN$\left(1\right)$    & 0.515(0.035)  \\  \hline
$3$-NN$_{ESOV}$               &  0.541(0.035)  &                          &               &                           &        \\  \hline \hline
\end{tabular}
\caption{Estimated rate of correct classification of the different approaches (with
standard errors in parentheses).}
\label{tab2}
\end{center}
\end{table}

\subsubsection*{Fatty acid and glass data from Examples 1 and 2} 

For both the fatty acid and forensic glass datasets, some of the groups have fewer
observations than the dimension $D$ of the compositions, so QDA cannot be applied (since
at least one of the $\hat{\bm{\Sigma}}_i$ in (\ref{rule}) is singular).  Both RDA, LDA
and $k$-NN are applicable, however, and Table \ref{tab1} shows a comparison of
performance for these techniques.

For the fatty acid data, RDA performs strongest, and best performance is achieved when
$\alpha=0.6$.  For this dataset $k$-NN($\alpha$) performs strongly too, with
$\alpha=0.35$ giving notably better performance than $\alpha=1$ (which corresponds to
the EDA approach).  For the forensic glass data, $k$-NN outperformed RDA and LDA, and
the flexibility of having $\alpha$ different from 1 offered no improvement.

For both of these datasets, results suggest that there is no clear relationship between
classification accuracy for the groups and the number of observations containing zeros,
i.e., no clear evidence that observations with zeros were more or less difficult to
classify correctly than those without zeros.  Figure \ref{figure2}, for example, shows
the classification accuracy for each group in the fatty acid dataset
plotted against the proportion of observations
that contain at least one zero, and no clear correlation is apparent.  
Results (not shown) for the
$k$-NN with the ESOV metric (\ref{ESOV}) similarly show little pattern.
Table \ref{tab2a} shows results for the fatty acid data presented according to the number
of zeros in the observations.  There is no clear relationsip between classification
accuracy and number of zeros.  Corresponding results in Table \ref{tab1b} for the forensic
glass data show lower classification accuracy for observations with 0 or 4 zeros compared
with observations with 1, 2 or 3 zeros, albeit with large standard errors on account of
the small number of such observations.  Hence, again, the conclusion is that there is no
clear evidence that zeros make observations any more or less difficult to classify
correctly.

The key points from these examples are that LRA is not directly applicable because of the
zeros, but EDA ($\alpha=1$) performs quite well with RDA having better performance in 
one example and $k$-NN in another, and in one of the examples letting $\alpha$ be a
value other than $1$ gave a further improvement.

\subsubsection*{Hydrochemical and national income data from Examples 3 and 4} 

For the hydrochemical data the extra flexibility of RDA over QDA offers no improvement
(and hence RDA and QDA give identical results).  Ill-conditioning of covariance matrices
makes QDA and LDA unstable for $\alpha>0.75$, which is why in Figure \ref{figure}(VII) the
lines corresponding to these methods stop at $\alpha=0.75$.  The plots in the left column
of Figure \ref{figure} show clearly that the performance of 
RDA$\left(\alpha,\lambda,\gamma \right)$ (and its special cases QDA($\alpha$) and
LDA($\alpha$)) depends on $\alpha$ and tend to do best at values of $\alpha$ other than 0
or 1. $k$-NN($\alpha$) does best for this example, with $\alpha=0.25$ and 2 nearest neighbours, leading to the best
performance of all the classifiers. 

For the final example of the national income data, the LRA approach of taking $\alpha=0$
leads to the best performance of RDA.  As in the previous example the $k$-NN classifier does best when $\alpha=-0.5$ and 2 neighbours are considered.

\section{Conclusions}

We have considered the $\alpha$-transformation (\ref{alpha}) and the $\alpha$-metric
(\ref{adist}) as a means to adapt LDA, QDA, RDA and $k$-NN for compositional data.  This
generalises EDA and LRA approaches via the parameter $\alpha$, the choice of which enable
a compromise between the two.  Rather than choosing either EDA or LRA, our approach
enables a choice of $\alpha$ based on the dataset at hand, and numerical results suggest
there is a clear benefit to having this flexibility.

An important benefit is that such an approach is well defined even when the dataset
contains observations with components equal to zero, unlike with LRA in which ad hoc
modifications to the data are needed prior to applying the log-ratio transformation.
Within $k$-NN it is simple to incorporate any choice of distance that seems appropriate.

\subsection*{Appendix}

\subsubsection*{Relationship between the $\alpha$-transformation and centred log-ratio transformation}
The proof that the transformation $(D \mathbf{u}_\alpha(\mathbf{x}) -
\mathbf{1}_D)/\alpha$ defined on the right-hand side of (\ref{alpha}) tends to the centred
log-ratio transformation (\ref{aitchisonclr}) as $\alpha \rightarrow 0$ is as follows:
for component $i$,
\begin{eqnarray*}
{\frac{1}{\alpha}\left(\frac{Dx_i^\alpha}{\sum_{j=1}^Dx_j^\alpha}-1\right)} &=&
\frac{D}{\alpha}\left[\frac{1+\alpha \log{x_i}+O\left(\alpha^2
\right)}{D\left(1+{\frac{\alpha}{D} \sum_{j=1}^D\log{x_j}}+O\left(\alpha^2 \right)\right)}-\frac{1}{D}\right] \\
&=&
\left\{{\left(1+\alpha\log{x_i}\right)\left(1+\frac{\alpha}{D}\sum_{j=1}^D\log{x_j}\right)^{-1}-1+O\left(\alpha^2
\right)}\right\}\Big\slash{\alpha}  \\
&=&\left\{\left(1+\alpha\log{x_i}\right)\left(1-\frac{\alpha}{D}\sum_{j=1}^D\log{x_j}\right)-1+O\left(\alpha^2
\right)\right\}\Big\slash{\alpha} \\
&=& \left\{1+\alpha\log{x_i}-\frac{\alpha}{D}\sum_{j=1}^D\log{x_j}-1+O\left( \alpha^2
\right)\right\}\Big\slash{\alpha} \\
&=& \log{x_i}-\log\prod_{j=1}^Dx_j^{1/D}+O\left( \alpha \right)  \\
&\rightarrow& \log\left\{\frac{x_i}{g\left( {\bf x} \right)}\right\} \ \ \text{as} \ \ \alpha \rightarrow 0. 
\end{eqnarray*}
The proof that the $\alpha$-metric (\ref{adist}) tends to the LRA metric
(\ref{aitchisondistance}) as $\alpha \rightarrow 0$ follows from this proof.


\end{document}